\documentclass[10pt]{article}
\usepackage[utf8]{inputenc}
\pdfoutput=1
\usepackage{natbib}
\usepackage[english]{babel}
\usepackage{hyperref,graphicx,xspace,float,afterpage}
\usepackage[pdftex]{color}
\usepackage[labelsep=period,labelfont=bf]{caption}
\usepackage{geometry}
\usepackage{latexsym,amsmath,amsfonts,amssymb,textcomp,calc,booktabs,url}
\DeclareMathOperator{\arccot}{\mathrm{arccot}}  
\newcommand{\df}{\textrm{d}}  
\newcommand{\dpow}[2]{\ensuremath{#1\cdot 10^{#2}}}  

\begin{document}
\title{Globally smooth approximations for shock pressure decay in impacts}
\author{Thomas Ruedas\thanks{Corresponding author: T. Ruedas, Institute of Planetology, Westf\"alische Wilhelms-Universit\"at, M\"unster, Germany (t.ruedas@uni-muenster.de)}\\{\footnotesize Institute of Planetology, Westf\"alische Wilhelms-Universit\"at, M\"unster, Germany}\\{\footnotesize Institute of Planetary Research, German Aerospace Center (DLR), Berlin, Germany}}
\date{}
\maketitle
\begin{abstract}
New forms of empirical formulae that provide an approximate description of the decay of shock pressure with distance in hypervelocity impacts are proposed. These forms, which are intended for use in applications such as large-scale mantle convection models, are continuous and smooth from the point of impact to arbitrarily large distances, thereby avoiding the need to divide the domain into different decay regimes and yielding the maximum pressure in a self-consistent way without resorting to the impedance-match solution. Individual fits for different impact velocities as well as a tentative general fitting formula are given, especially for the case of dunite-on-dunite impacts. The temperature effects resulting from the shock are estimated for different decay models, and the differences between them are found to be substantial in some cases, potentially leading to over- or underestimates of impact heating and melt production in modeling contexts like mantle convection, where such parameterizations are commonly used to represent giant impacts.
\end{abstract}

\begin{flushleft}
Impact processes; Terrestrial planets; Thermal histories; Interiors
\end{flushleft}

\section{Introduction}
Hypervelocity impacts on a planet subject the target material to extreme conditions, in particular very high pressures. It has long been known from experimental and numerical studies that the impactor penetrates the target before having deposited all of its energy, so that the center of the region from which the shock wave emerges lies at a certain depth beneath the pre-impact surface. The pressure of the shock varies with some inverse power of distance from this center, but it does not do so in a uniform way. Rather, the finite size of the projectile causes pressure in a target region comparable to its own size to vary in a different way than at greater distance \citep[e.g.,][]{GaHe63}, and the details of this variation are controlled by the shape of the impactor. In the context of planetary science, the most common configuration considered is that of a spherical impactor hitting a homogeneous half-space. Since the pioneering numerical study by \citet{AhOKe77}, which investigated this type of configuration, expressions of the form
\begin{equation}
\lg p=\tilde a+\tilde n\lg\left(\frac{r}{R}\right)\label{eq:tradform}
\end{equation}
have been used to describe the decay of shock pressure $p$ with distance $r$; $R$ is the radius of the impactor, $\tilde a$ and $\tilde n$ are fitting constants, and lg is the decadic logarithm. The change in the character of pressure decay implies that the expression eq.~\ref{eq:tradform} does not apply with the same $\tilde a$ and $\tilde n$ everywhere, but that piecewise fits must be carried out in the different domains. The numerical studies by \citet{AhOKe77}, \citet{Pier:etal97}, and \citet{MoAr-Ha16} indicate that the pressure drop is much smaller at small $r$ than at great distance from the impact site. Most studies on the subject distinguish between a near-field and a far-field and perform fits for these two domains; only \citet{MoAr-Ha16} have chosen to introduce an additional narrow mid-field. Such expressions for the decay of the impact shock can be used to estimate the heating of the target following the formalism developed by \citet{GaHe63} or a variant thereof \citep{Melosh89}, and they have been used to define the thermal perturbations $\Delta T$ caused by impacts in planetary bodies that influence the course of evolution of mantle (and maybe even core) convection \citep[e.g.,][]{Rees:etal02,WAWatt:etal09,JHRoAr-Ha12}. Global dynamical evolution models of this type often do not require the highly detailed representation of impacts provided by computationally expensive fully dynamical impact simulations, and their substitution with simple and readily computable first-order representations such as those developed in the following are the principal motivation for this paper.\par
The power-law form of eq.~\ref{eq:tradform} has been found to work well over limited ranges and stands in a tradition of many other empirical or semi-empirical relations in impact studies that have the form of a power law, and its simplicity makes it easy to implement. However, in this case it also has disadvantages. Obviously, the term $\lg(r/R)$ goes to infinity as $r\to 0$, which means that an upper bound of $p$ has to be imposed in some form; the application of the Hugoniot equation of state to the idealized situation of two colliding infinite planes offers itself as the most straightforward and familiar analytical model if such a bound is not derived directly from the data. Indeed, the near-field region, in which pressure decay with distance is weak, has been treated as isobaric by several workers, and the pressure within this isobaric core has been determined from that so-called impedance-match solution \citep[e.g.,][]{WAWatt:etal09}. While this is probably the best they could do with the available means, it implies that the impedance-match solution is indeed the correct upper bound for $p$ and also an acceptable approximation for the supposedly constant value of $p$ in the near field or ``isobaric core'', but this is not always supported, if not directly contradicted, by numerical models. With this notion of a (quasi-)isobaric central region in mind, \citet{Pier:etal97} even derived an empirical formula for the radius of the isobaric core from their numerical models. The concept of this bimodality of pressure decay is a helpful qualitative description, but on the other hand the division into distinct domains introduces discontinuities in the radial derivatives of $p$ and $\Delta T$ that are unnatural. These remarks on the shock pressure apply in similar form also to the particle velocity, which is closely related.\par
In this paper, alternative formulations for the pressure decay with distance are considered. These formulations are smooth (in the sense of having a first derivative) everywhere, finite, and applicable with a single set of two or three parameters for all $r$. Some consideration is also given to the possibility of defining domains such as a ``near field'' and a ``far field'' self-consistently from properties of the fitting model. Due to the somewhat tentative character of the results of this paper and because the pressure is probably of more immediate interest for work outside the impact community than the particle velocity, the focus here is exclusively on the pressure. Although formulae of the type discussed in this paper have their merits as a fairly easy and computationally inexpensive means of estimating the most important first-order effects of large impacts and thus allow their straightforward inclusion in global convection models, it is important to keep in mind that the assumptions on which they are based, i.e., the combination of the (one-dimensional) impedance-match solution for the pressure with a radially symmetric decay profile, are very idealized. Therefore, they are not an accurate method of describing the smaller-scale details of any given impact process or the deviations from radial symmetry; for that purpose, further assumptions would be needed.

\section{Results}
In order to avoid the shortcomings of the model function eq.~\ref{eq:tradform}, an alternative formulation for $p(r)$ should be 1) positive for all finite $r$ and bounded from above; 2) monotonically decreasing to zero for $r\geq 0$; 3) almost constant at $r\lesssim R$; and 4) asymptotically equal to the form of eq.~\ref{eq:tradform} for large $r$. Requirements 2 and 3 suggest that $p(r)$ may have a maximum at $r=0$, i.e., $\mathrm{d}p/\mathrm{d}r=0$; this may not necessarily be the case, although it would be more consistent with the idealized impedance-match solution for infinite colliding planes, which is briefly summarized in \ref{app:im}. The physics involved in an impact is too complicated to be captured in a single equation, and it is likely that different processes dominate at different distances from the impact center, which entail different decay behaviors for $p(r)$; a change in the exponent of $r/R$ had already been pointed out by \citet{GaHe63}. Nonetheless, the four requirements provide a framework in which a heuristic mathematical approximation for easy use in applications that do not demand highest precision can be sought. There are various functions that fulfill these conditions; here I investigate the three forms,
\begin{subequations}\label{eqs:genform}
\begin{align}
p&=\frac{a}{b+\left(\frac{r}{R}\right)^n}\\
p&=\frac{a}{\left(b+\frac{r}{R}\right)^n}\\
p&=a\arccot\left[b\left(\frac{r}{R}\right)^n\right],
\end{align}
\end{subequations}
which will hereafter be referred to as the ``inverse-$r$'', the ``inverse polynomial'', and the ``arc cotangent'' model functions, respectively. Other forms, especially generalized Gaussian bell functions, have also been considered, but they always seem to decay too strongly with $r$. The coefficients $a$ and $b$ in the first two forms are coupled by the fact that $p(0)=a/b$ and $p(0)=a/b^n$, respectively. The impedance-match approximation is expected to be nowhere as accurate as at $r=0$, and it provides a natural scale for the shock pressure. It is convenient to determine the fit in terms of a normalized pressure $p/p_\mathrm{IM}$, where $p_\mathrm{IM}$ is the impedance-match solution. Indeed, if one decides to impose the impedance-match solution at $r=0$ as a constraint, i.e., $p(0)/p_\mathrm{IM}=1$, then the number of free parameters is reduced by one as $a=b$, $a=b^n$, and $a=2/\pi$, respectively, and the model equations become
\begin{subequations}\label{eqs:cform}
\begin{align}
\frac{p}{p_\mathrm{IM}}&=\frac{b'}{b'+\left(\frac{r}{R}\right)^{n'}}\\
\frac{p}{p_\mathrm{IM}}&=\left(\frac{b'}{b'+\frac{r}{R}}\right)^{n'}\\
\frac{p}{p_\mathrm{IM}}&=\frac{2}{\pi}\arccot\left[b'\left(\frac{r}{R}\right)^{n'}\right].
\end{align}
\end{subequations}
Both the general forms eqs.~\ref{eqs:genform} (normalized with $p_\mathrm{IM}$) and the ``constrained'' forms eqs.~\ref{eqs:cform} have been applied to various datasets for which both the numerical data and the parameters necessary to determine the impedance-match solution were available. The most important ones are those for dunite projectiles and targets from \citet{Pier:etal97} for impactor velocities $v$ between 10 and 60\,km/s and from \citet{MoAr-Ha16} for $v$ between 4 and 10\,km/s, which match up quite well at 10\,km/s and can therefore be studied separately as well as in combination; although \citet{MoAr-Ha16} do not give all material properties in the same explicit form as \citet{Pier:etal97}, their discussion and the good match between both datasets justifies to use the same properties for both. In addition, the data for gabbroic anorthosite or iron projectiles and gabbroic anorthosite targets from \citet{AhOKe77} were used in order to test the applicability to different materials and to include a case in which impactor and target consist of different materials, although their datasets are sparser and of lower quality. All data points were extracted from figures in the respective publications and are shown in normalized form in Figure~\ref{fig:data}. The material properties and the resulting $p_\mathrm{IM}$ are given in Table~\ref{tab:matprop}.\par
\begin{figure}[b]
\captionof{figure}{$p(r) /p_\mathrm{IM}$ for different impactor velocities $v$ and different materials from numerical models by \citet{Pier:etal97} ($v\geq 10$\,km/s, open symbols), \citet{MoAr-Ha16} ($v\leq 10$\,km/s, solid symbols) (a) and \citet{AhOKe77} (b and c). The distance $r$ is normalized with the impactor radius $R$, the shock pressures $p$ are normalized with the corresponding impedance-match solutions $p_\mathrm{IM}$.\label{fig:data}}
\end{figure}
\begin{figure}[!p]
\begin{center}
\includegraphics[width=0.83\textwidth]{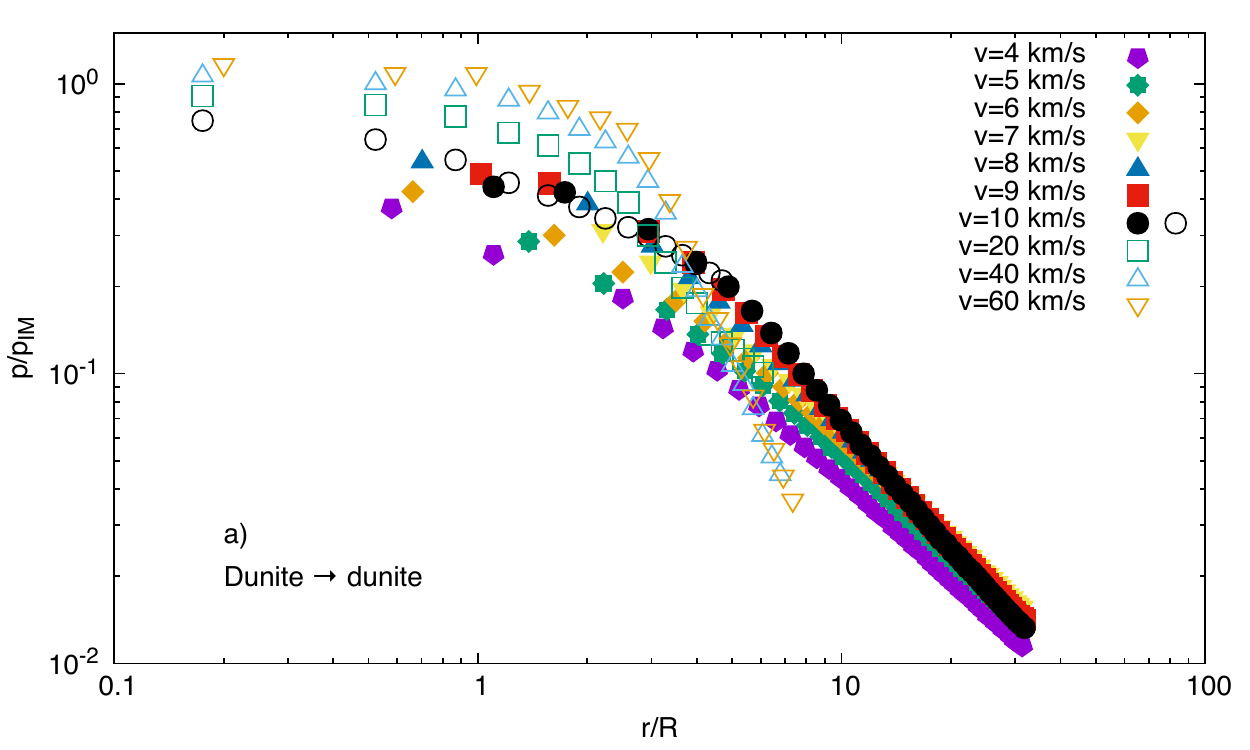}
\includegraphics[width=0.83\textwidth]{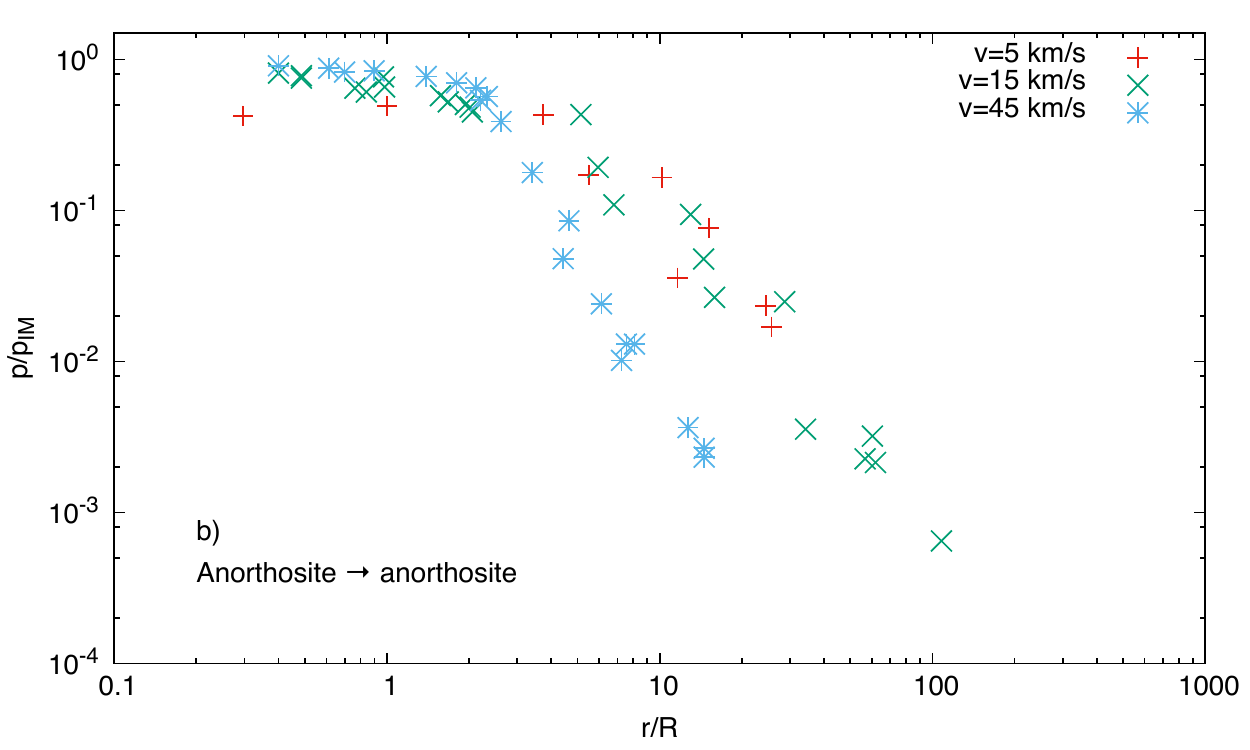}
\includegraphics[width=0.83\textwidth]{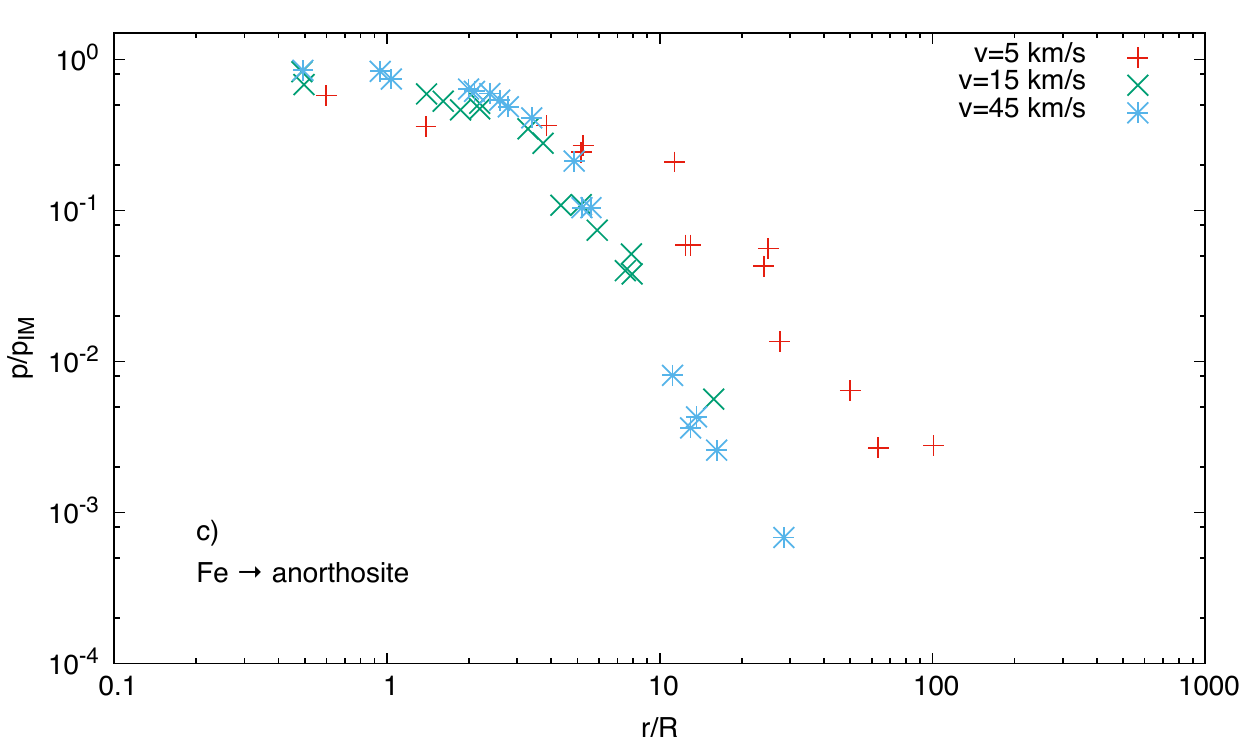}
\end{center}
\end{figure}
\begin{table}
\caption{Density $\varrho$, bulk sound speed $c$, slope $S$ of the Hugoniot curve in shock front velocity--particle velocity coordinates, and the resulting impedance match shock pressure $p_\mathrm{IM}$. The dunite data are from \citet[Tab.~AII.2]{Melosh89} and \citet[Tab.~1]{Pier:etal97}, the other data were calculated from the Tillotson equation of state parameters for iron and gabbroic anorthosite (low-pressure phase) as given by \citet[Tab.~1]{AhOKe77} (cf. eq.~\ref{eq:tilleos}).\label{tab:matprop}}
\begin{center}
\begin{tabular}{lc}\toprule
$v$ (km/s)&$p_\mathrm{IM}$ (GPa)\\\midrule
\multicolumn{2}{l}{Dunite $\to$ dunite ($\varrho=3320$\,kg/m\textsuperscript{3}, $c=6.5$\,km/s, $S=0.9$)}\\
4&55.112\\
5&72.625\\
6&91.632\\
7&112.133\\
8&134.128\\
9&157.617\\
10&182.6\\
20&514.6\\
40&1626.8\\
60&3336.6\\
\multicolumn{2}{l}{Anorthosite $\to$ anorthosite ($\varrho=2936$\,kg/m\textsuperscript{3}, $c=4.9$\,km/s, $S=1.194$)}\\
5&57.875\\
15&305.071\\
45&2098.224\\
\multicolumn{2}{l}{Fe $\to$ anorthosite (Fe: $\varrho=7800$\,kg/m\textsuperscript{3}, $c=4.049$\,km/s, $S=1.41$)}\\
5&85.593\\
15&471.895\\
45&3349.465\\\bottomrule
\end{tabular}
\end{center}
\end{table}
The fitting procedure was carried out with a Levenberg--Marquardt algorithm and consisted of two steps. First, each dataset for a given $v$ was fitted to each of the six model functions as a function of the form $p(r)/p_\mathrm{IM}=f(r/R)$ to determine the fitting parameters for each $v$; for the dunite/dunite data and $v=10$\,km/s, where two complementary datasets were available, both sets were fitted individually as well as in combination. The misfit was calculated as the root-mean-square of the residuals,
\begin{equation}
\sigma_p=\sqrt{\frac{1}{N-\nu}\sum^N (p_\mathrm{s,data}-p_\mathrm{s,regr.})^2},
\end{equation}
where $N-\nu$ is the number of degrees of freedom of the fit \citep[cf.][]{JRTaylor97}; $N$ is the number of data points and $\nu$ is the number of fitting parameters, i.e., 3 for eqs.~\ref{eqs:genform} and 2 for the constrained forms. The results of selected
$p$ fits are given in Table~\ref{tab:fit1}, a complete table is provided as Supplementary Material. Fig.~\ref{fig:fit1} shows the data and some fits for dunite and impactor velocities of 4, 10, and 60\,km/s. Also shown are the fitting formulae from \citet{MoAr-Ha16} for the corresponding velocities (or the general formula with $v$-dependent coefficients for the 60\,km/s-model) and the formula from \citet{Pier:etal97}; as the latter does not provide a scaling factor for the pressure, I followed what seems to be common practice in the use of this formula, namely to determine the radius of the isobaric core with their formula (using their values for dunite) and to set $p$ to the impedance-match solution at $r\leq r_\mathrm{ic}$ while applying their decay law scaled to give a continuous graph outside. Furthermore, a modified version of their formula was also tried out as a consequence of certain problems with their original formula that will be discussed below.\par
\afterpage{\clearpage}
\begin{table}[!p]
\caption{Selected results of fit to eqs.~\ref{eqs:genform} and \ref{eqs:cform}, rounded to three decimal places. In the dunite fits for 10\,km/s, the individual fits to the data from \citet{MoAr-Ha16} and \citet{Pier:etal97}, respectively, and the combined fit are given. Full results are compiled in Table~1 of the Supplementary Material.\label{tab:fit1}}
\begin{center}
\begin{tabular}{cccccccc}\toprule
$v$ (km/s)&$a$&$b$&$n$&misfit&$b'$&$n'$&misfit\\\midrule
\multicolumn{8}{l}{Inverse-$r$ model (eqs.~\ref{eqs:genform}a and \ref{eqs:cform}a)}\\
\multicolumn{8}{l}{Dunite $\to$ dunite}\\
4&0.797&1.687&1.213&\dpow{4.957}{-3}&0.392&0.904&\dpow{9.384}{-3}\\
5&1.148&2.551&1.285&\dpow{2.069}{-3}&0.598&1.031&\dpow{5.291}{-3}\\
6&1.367&2.645&1.307&\dpow{1.262}{-3}&0.591&0.945&\dpow{1.217}{-2}\\
7&1.355&1.575&1.299&\dpow{6.607}{-4}&1.236&1.265&\dpow{9.121}{-4}\\
8&2.899&4.710&1.590&\dpow{3.845}{-3}&1.062&1.123&\dpow{1.810}{-2}\\
9&4.475&7.934&1.747&\dpow{3.896}{-3}&1.292&1.198&\dpow{1.736}{-2}\\
10&6.489&13.111&1.883&\dpow{4.285}{-3}&1.231&1.165&\dpow{2.119}{-2}\\
&1.311&1.564&0.994&\dpow{8.126}{-3}&0.976&0.785&\dpow{1.800}{-2}\\
&1.892&2.566&1.317&\dpow{1.771}{-2}&1.074&1.023&\dpow{2.995}{-2}\\
20&5.224&5.888&2.243&\dpow{1.820}{-2}&3.191&1.861&\dpow{3.712}{-2}\\
40&18.661&18.352&3.040&\dpow{3.096}{-2}&20.924&3.124&\dpow{3.074}{-2}\\
60&34.451&31.635&3.366&\dpow{3.915}{-2}&70.989&3.884&\dpow{5.234}{-2}\\
\multicolumn{8}{l}{Anorthosite $\to$ anorthosite}\\
5&37.691&80.474&2.447&\dpow{6.207}{-2}&0.616&0.551&\dpow{1.147}{-1}\\
15&3.109&3.720&1.340&\dpow{5.756}{-2}&1.850&1.058&\dpow{6.090}{-2}\\
45&57.067&66.452&4.439&\dpow{2.756}{-2}&12.287&3.007&\dpow{6.654}{-2}\\
\multicolumn{8}{l}{Fe $\to$ anorthosite}\\
5&5.250&9.761&1.440&\dpow{5.307}{-2}&0.973&0.775&\dpow{5.987}{-2}\\
15&7.149&9.489&2.335&\dpow{5.088}{-2}&2.144&1.461&\dpow{7.567}{-2}\\
45&26.179&31.872&2.996&\dpow{2.912}{-2}&6.513&2.000&\dpow{6.099}{-2}\\
\multicolumn{8}{l}{Arc cotangent model (eqs.~\ref{eqs:genform}c and \ref{eqs:cform}c)}\\
\multicolumn{8}{l}{Dunite $\to$ dunite}\\
4&0.280&0.539&1.069&\dpow{5.925}{-3}&2.155&0.796&\dpow{1.140}{-2}\\
5&0.253&0.344&1.146&\dpow{2.495}{-3}&1.433&0.926&\dpow{6.860}{-3}\\
6&0.317&0.395&1.129&\dpow{2.633}{-3}&1.528&0.816&\dpow{1.519}{-2}\\
7&0.388&0.374&1.217&\dpow{5.934}{-4}&0.745&1.134&\dpow{2.563}{-3}\\
8&0.385&0.250&1.367&\dpow{1.474}{-3}&0.942&0.953&\dpow{2.184}{-2}\\
9&0.351&0.158&1.502&\dpow{2.236}{-3}&0.795&1.020&\dpow{2.078}{-2}\\
10&0.311&0.108&1.601&\dpow{2.963}{-3}&0.824&0.994&\dpow{2.449}{-2}\\
&0.543&0.727&0.778&\dpow{7.937}{-3}&1.015&0.631&\dpow{1.760}{-2}\\
&0.461&0.434&1.111&\dpow{2.083}{-2}&0.928&0.859&\dpow{3.390}{-2}\\
20&0.566&0.234&1.834&\dpow{1.738}{-2}&0.385&1.521&\dpow{3.637}{-2}\\
40&0.647&0.090&2.511&\dpow{3.173}{-2}&0.080&2.582&\dpow{3.147}{-2}\\
60&0.692&0.055&2.804&\dpow{3.931}{-2}&0.028&3.241&\dpow{5.213}{-2}\\
\multicolumn{8}{l}{Anorthosite $\to$ anorthosite}\\
5&0.300&0.028&2.006&\dpow{6.201}{-2}&1.492&0.447&\dpow{1.176}{-1}\\
15&0.535&0.349&1.093&\dpow{5.886}{-2}&0.606&0.867&\dpow{6.277}{-2}\\
45&0.548&0.032&3.654&\dpow{2.636}{-2}&0.126&2.483&\dpow{6.469}{-2}\\
\multicolumn{8}{l}{Fe $\to$ anorthosite}\\
5&0.336&0.132&1.244&\dpow{5.352}{-2}&1.025&0.641&\dpow{6.285}{-2}\\
15&0.480&0.159&1.909&\dpow{5.216}{-2}&0.522&1.210&\dpow{7.721}{-2}\\
45&0.525&0.061&2.433&\dpow{2.985}{-2}&0.209&1.659&\dpow{6.075}{-2}\\\bottomrule
\end{tabular}
\end{center}
\end{table}
\begin{figure}[!b]
\captionof{figure}{Numerical model data for dunite and different fits of $p(r)/p_\mathrm{IM}$ for impactor velocities $v$ of 4 (a), 10 (b), and 60\,km/s (c); the data points are from \citet{Pier:etal97} (PVM97) and \citet{MoAr-Ha16} (MAH16). Note that the PVM97 and MAH16 formulae are used outside their range of calibration in panels (a) and (c), respectively, which explains their unphysical behavior there. A version of this plot that includes only the inverse-$r$ and the arc cotangent curves and provides error bars can be found on pp.~27--28 of the Supplementary Material.\label{fig:fit1}}
\end{figure}
\begin{figure}[!p]
\begin{center}
\includegraphics[width=0.83\textwidth]{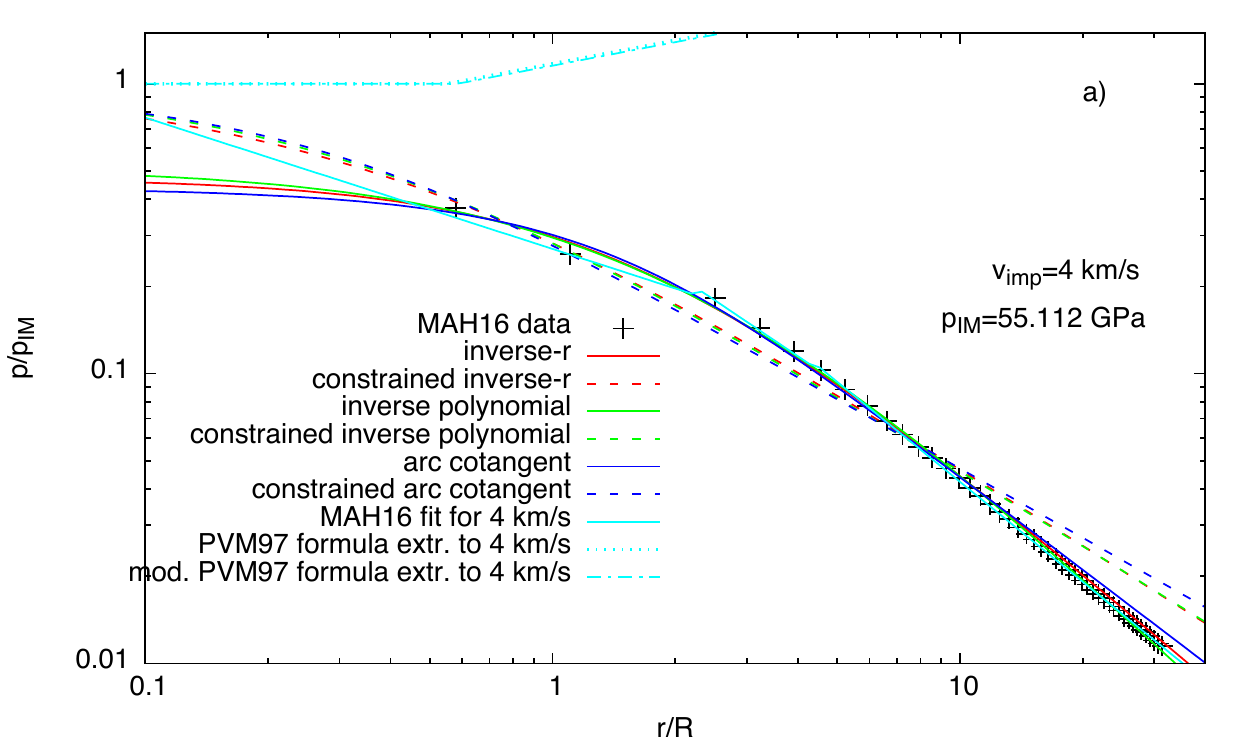}
\includegraphics[width=0.83\textwidth]{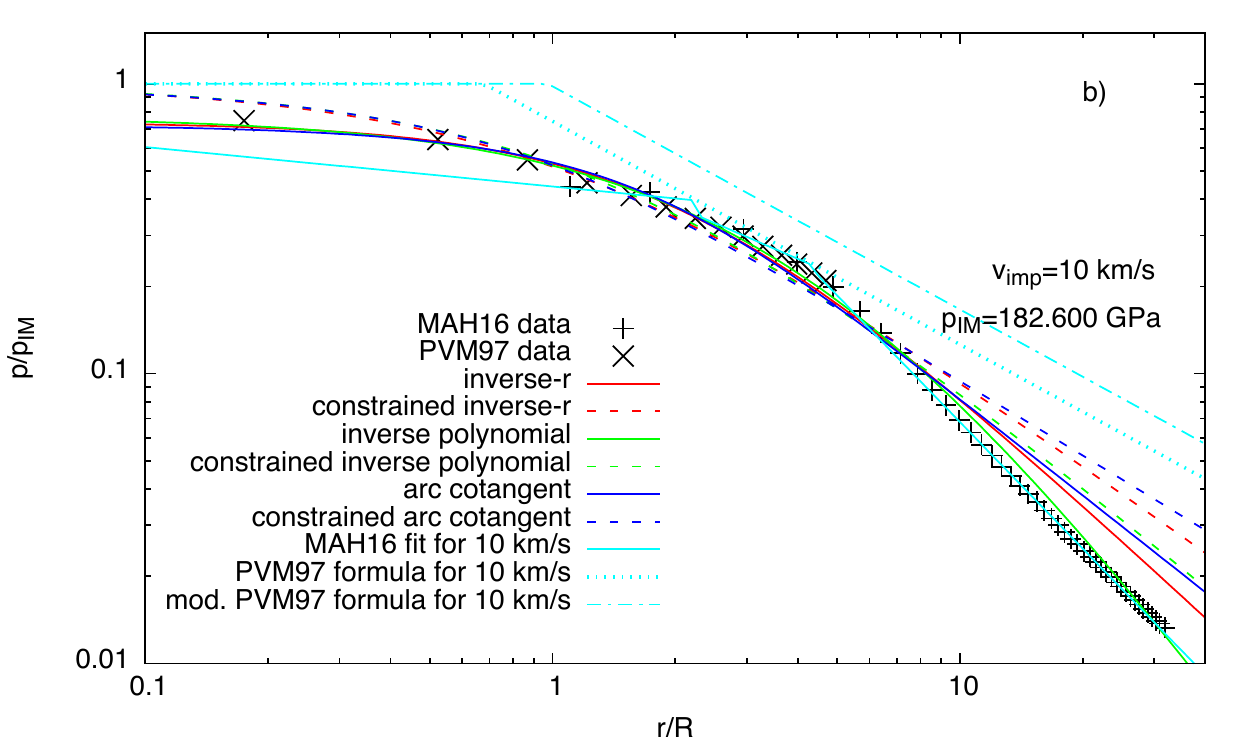}
\includegraphics[width=0.83\textwidth]{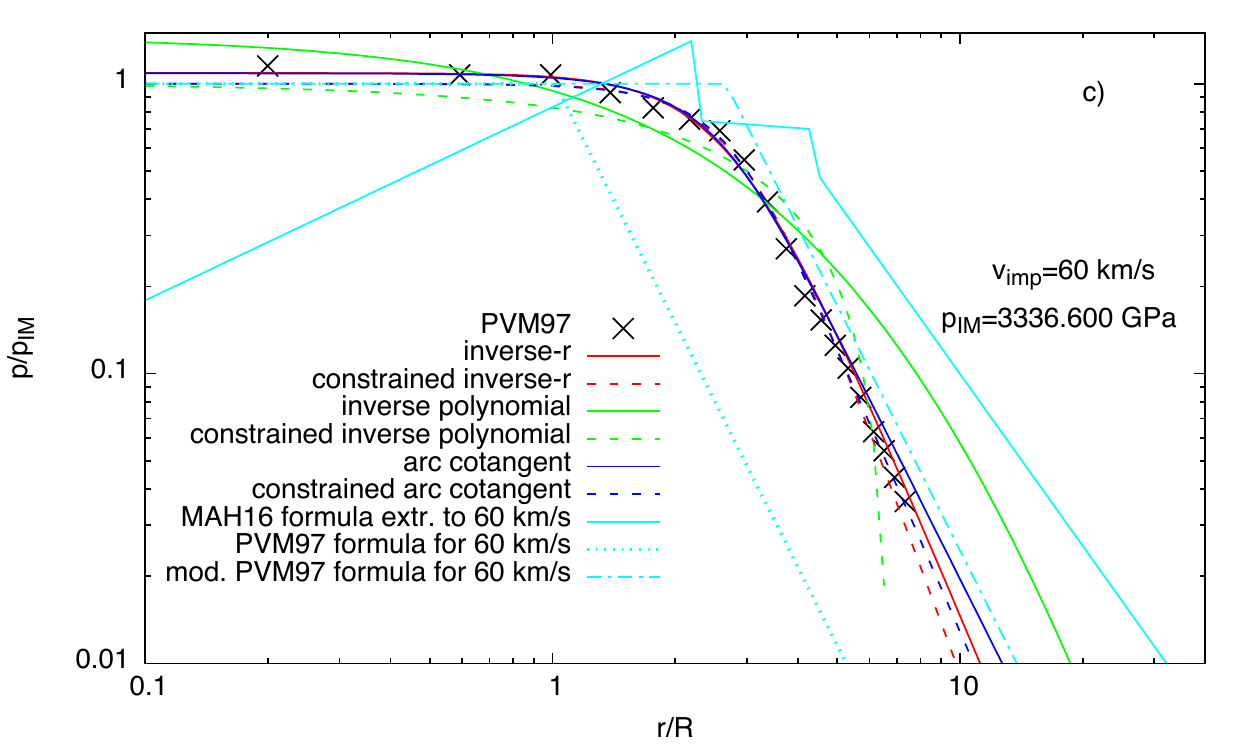}
\end{center}
\end{figure}
At least some of the fitting parameters $a$, $b$, and $n$ are themselves found to be functions of impactor velocity $v$, in agreement with the findings from the earlier studies, although the type of dependence is not always clear and straightforward to cast in a simple form; it may even depend on the materials involved. Two different types of functions are tried, namely linear functions of the form
\begin{subequations}\label{eqs:parfit}
\begin{equation}
f(v)=\alpha+\beta v
\end{equation}
for all three parameters and power-law type functions of the form
\begin{equation}
f(v)=\alpha v^\beta
\end{equation}
for $a$ and $b$; in the case of $n$, the suggestion by \citet{AhOKe87} to use logarithmic model functions of the form
\begin{equation}
n(v)=\alpha+\beta\lg v
\end{equation}
\end{subequations}
seems to be a more appropriate choice for the decay functions eqs.~\ref{eqs:genform} and \ref{eqs:cform} as well, again in agreement with the earlier work. Hence, the second step was to try to fit the parameters $a$, $b$, and $n$ to each of the model functions eqs.~\ref{eqs:parfit} for each set of materials; again, the 10\,km/s data for dunite were treated individually as well as in combination. Moreover, the 20\,km/s dunite data were excluded from the parameter fit, because especially the values for $a$ and $n$ are so far off the general trend that they should be regarded as outliers. In some cases, however, the data in this second step have such a markedly non-monotonic trend or show so little variation that no convincing fit can be achieved and that the model function is dismissed on these grounds, or a mean value should be used instead. The results of selected parameter fits of interest carried out are shown in Fig.~\ref{fig:parfit} and given in Table~\ref{tab:fit2}, the complete data are provided as Supplementary Material.\par
\begin{figure}[!p]
\begin{center}
\includegraphics[width=0.75\textwidth]{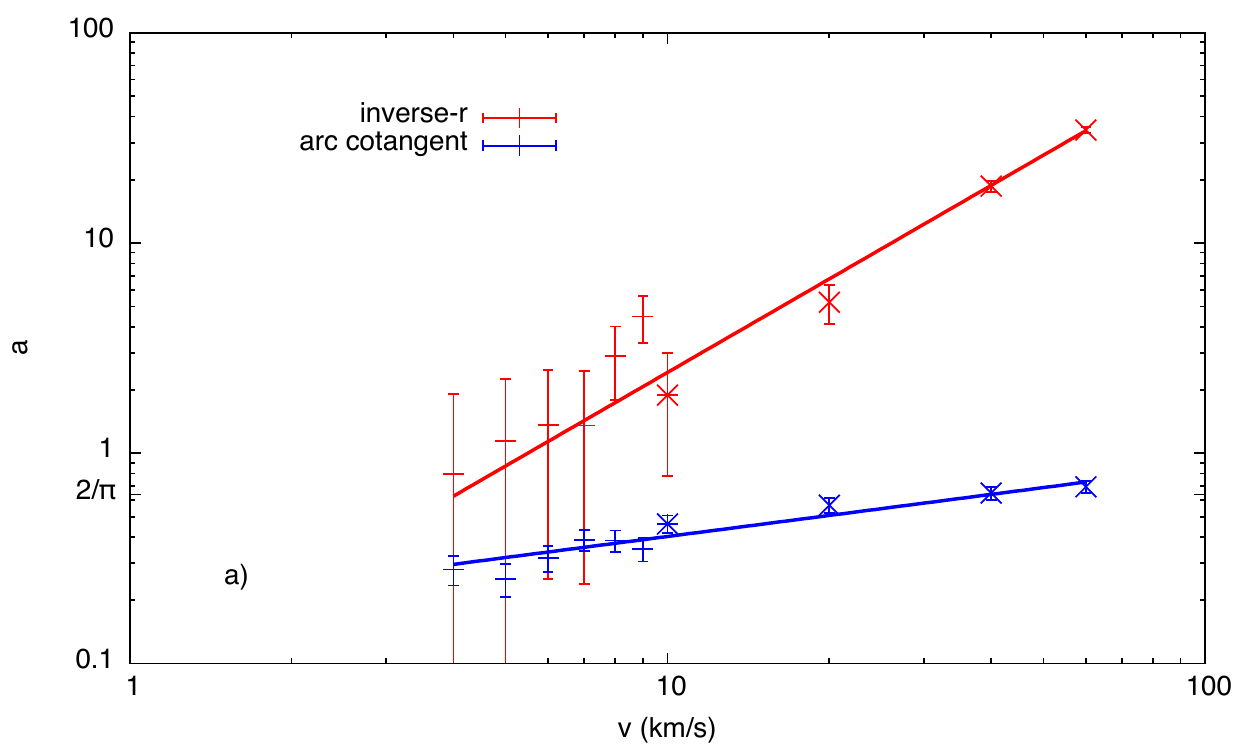}
\includegraphics[width=0.75\textwidth]{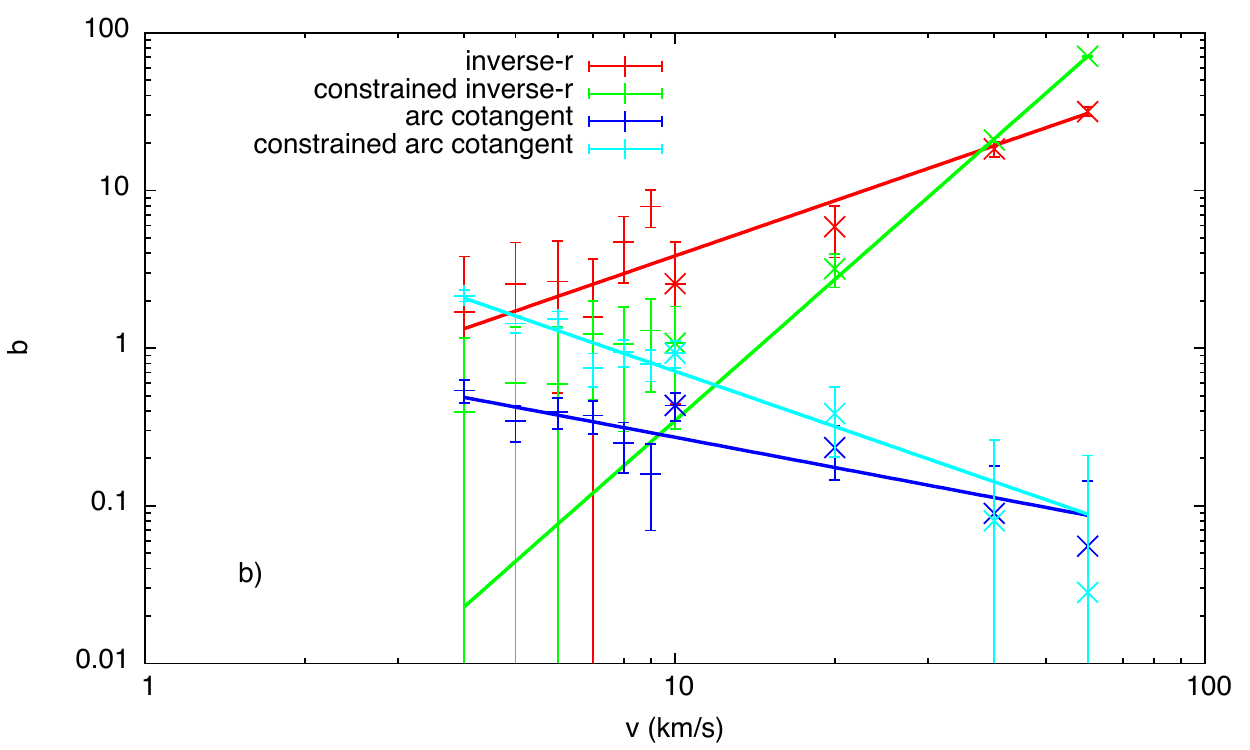}
\includegraphics[width=0.75\textwidth]{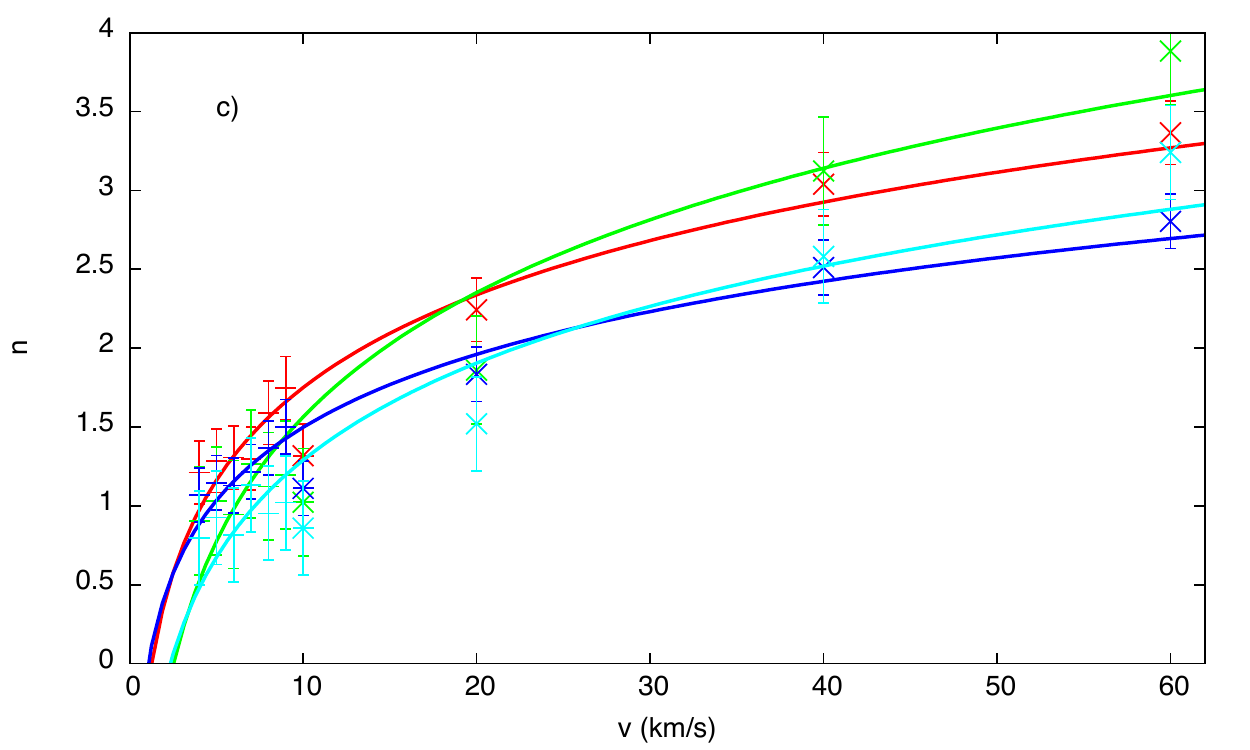}
\end{center}
\caption{Fits of $a$ (a), $b$ (b), and $n$ (c) as a function of impactor velocity $v$, for dunite and using the preferred power law model eq.~\ref{eqs:parfit}b in the case of $a$ and $b$ and the logarithmic function model eq.~\ref{eqs:parfit}c in the case of $n$. The data points and their error bars represent the parameter values for a given $v$ and their standard deviations, based on models by \citet{MoAr-Ha16} ($+$) and \citet{Pier:etal97} ($\times$); the data points for $v=10$\,km/s are from the combined fit to both datasets ($+\llap{$\times$}$). The corresponding fits to the linear model eq.~\ref{eqs:parfit}a are shown on page 31--32 of the Supplementary Material.\label{fig:parfit}}
\end{figure}
\begin{figure}[!p]
\begin{center}
\includegraphics[width=0.75\textwidth]{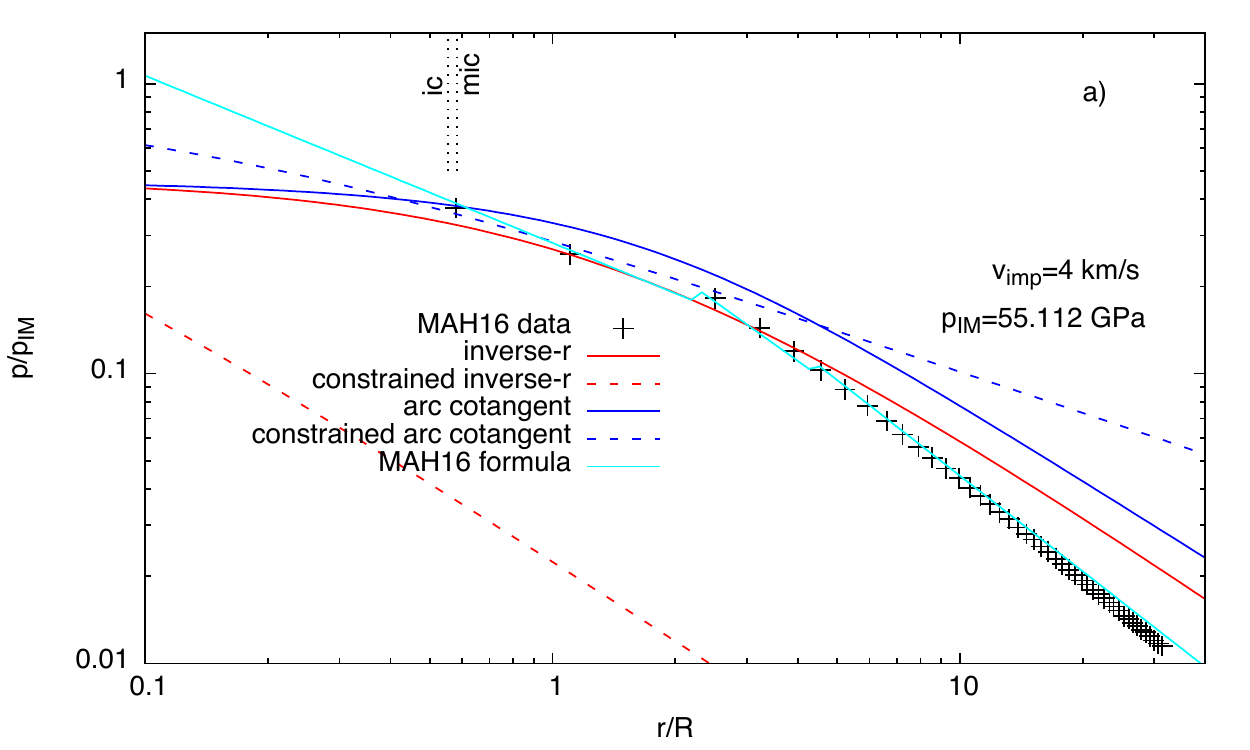}
\includegraphics[width=0.75\textwidth]{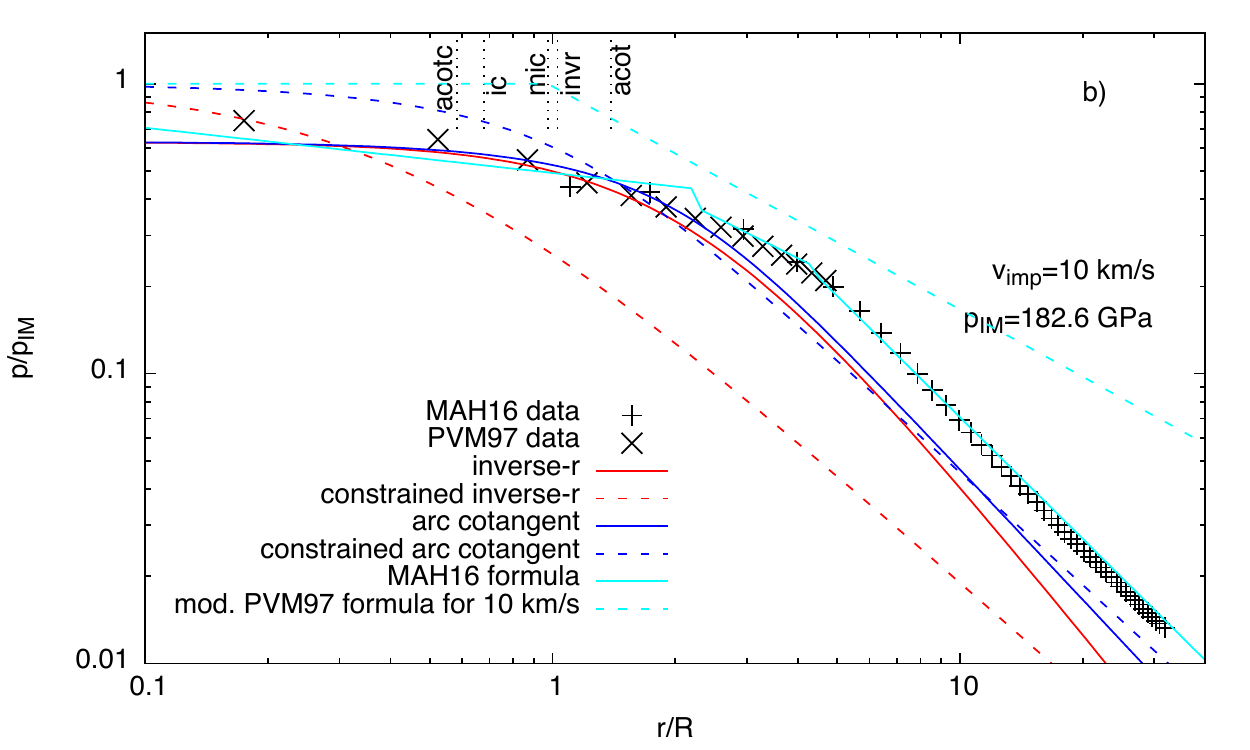}
\includegraphics[width=0.75\textwidth]{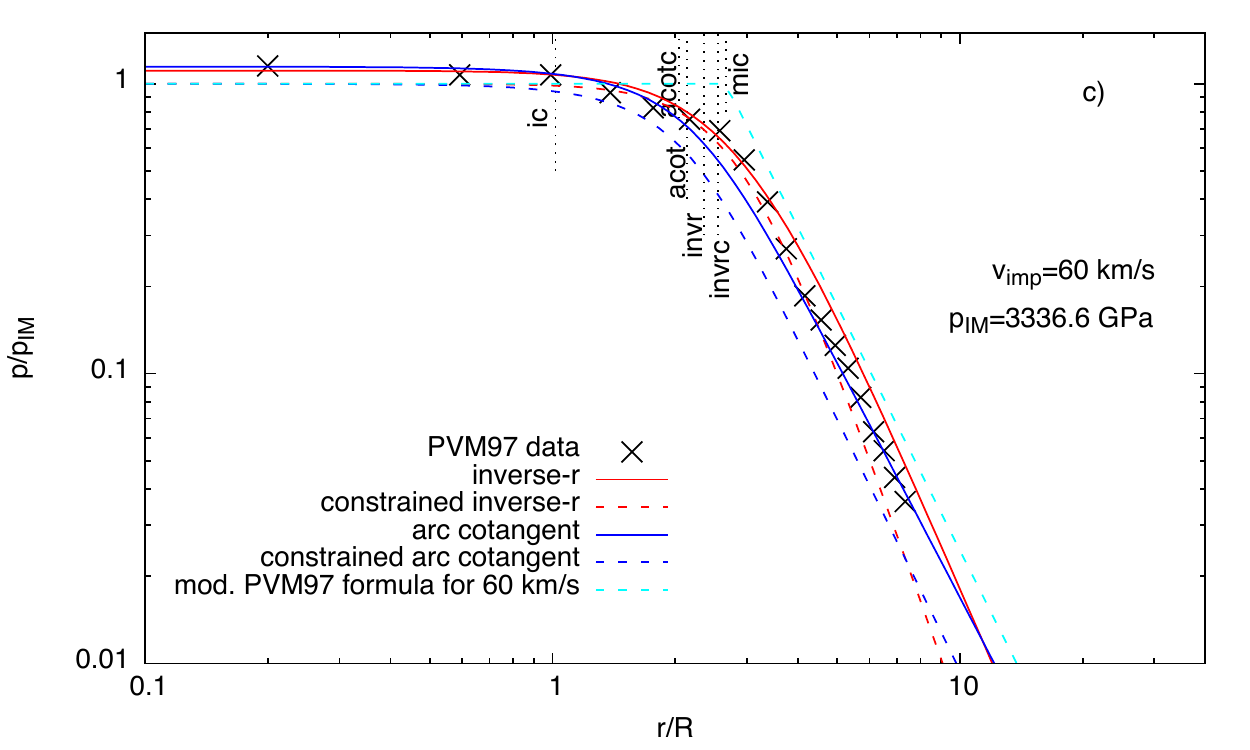}
\end{center}
\caption{Same as Fig.~\ref{fig:fit1} with the general fitting formulae, for the inverse-$r$ and arc cotangent decay models; a version of this plot with error bars is provided on pp.~29--30 of the Supplementary Material. The constrained inverse-$r$ curve used a linear $n(v)$, the other a logarithmic one. The dotted vertical lines with labels indicate the positions of the inflexion points of the fitting curves where applicable (invr: inverse-$r$; invrc: constrained inverse-$r$; acot: arc cotangent; acotc: constrained arc cotangent) and of the radius of the isobaric core for dunite according to \citet{Pier:etal97} (ic: original; mic: modified).\label{fig:fit1g}}
\end{figure}

\begin{table}[!p]
\caption{Selected results of fits to eqs.~\ref{eqs:parfit}, rounded to three decimal places. The three lines in the dunite datasets correspond to the data from \citet{MoAr-Ha16}, \citet{Pier:etal97}, and the combined fits, respectively. ``Fe/An'' and ``Anorth.'' correspond to the models from \citet{AhOKe77} for iron and anorthosite projectiles, respectively, hitting an anorthosite target. Full results are compiled in Table~2 of the Supplementary Material.\label{tab:fit2}}
\begin{center}
\begin{tabular}{lcccccc}\toprule
&\multicolumn{3}{c}{linear fit (eq. \ref{eqs:parfit}a)}&\multicolumn{3}{c}{power law fit (eq. \ref{eqs:parfit}b)}\\
Data&$\alpha$&$\beta$&misfit&$\alpha$&$\beta$&misfit\\\midrule
\multicolumn{7}{c}{$a$}\\
\multicolumn{7}{l}{Inverse-$r$}\\
Dunite&$-3.668$&0.902&\dpow{9.328}{-1}&0.003&3.336&\dpow{4.309}{-1}\\
&$-6.958$&0.673&1.837&0.044&1.631&\dpow{7.894}{-1}\\
&$-2.569$&0.580&1.972&0.081&1.478&1.115\\
Fe/An&0.967&0.549&2.615&0.592&0.993&2.811\\
\multicolumn{6}{l}{Arc cotangent}\\
Dunite&0.237&0.013&\dpow{4.724}{-2}&0.192&0.278&\dpow{4.539}{-2}\\
&0.511&0.003&\dpow{1.054}{-2}&0.381&0.144&\dpow{1.732}{-2}\\
&0.308&0.007&\dpow{6.941}{-2}&0.187&0.333&\dpow{4.500}{-2}\\
Anorth.&0.355&0.005&\dpow{1.355}{-1}&0.245&0.226&\dpow{1.029}{-1}\\
Fe/An&0.361&0.004&\dpow{7.545}{-2}&0.267&0.186&\dpow{4.779}{-2}\\\midrule
\multicolumn{7}{c}{$b$}\\
\multicolumn{7}{l}{Inverse-$r$}\\
Dunite&$-6.889$&1.682&2.425&0.001&4.061&1.316\\
&$-5.445$&0.609&1.042&0.075&1.477&\dpow{8.375}{-1}\\
&$-0.655$&0.509&2.284&0.266&1.161&2.123\\
Fe/An&4.097&0.597&4.550&1.542&0.789&5.625\\
\multicolumn{7}{l}{Constrained inverse-$r$}\\
Dunite&$-0.179$&0.156&\dpow{1.797}{-1}&0.097&1.151&\dpow{1.837}{-1}\\
&$-21.090$&1.388&13.018&\dpow{3.489}{-4}&2.985&\dpow{6.044}{-1}\\
&$-8.827$&1.122&8.256&\dpow{3.747}{-4}&2.968&\dpow{7.660}{-1}\\
Anorth.&$-1.685$&0.305&1.321&0.021&1.676&\dpow{3.223}{-1}\\
Fe/An&0.174&0.140&\dpow{1.674}{-1}&0.170&0.956&\dpow{2.197}{-1}\\
\multicolumn{7}{l}{Arc cotangent}\\
Dunite&0.761&$-0.065$&\dpow{5.556}{-2}&2.826&$-1.195$&\dpow{7.195}{-2}\\
&0.653&$-0.012$&\dpow{2.126}{-1}&26.737&$-1.567$&\dpow{1.218}{-2}\\
&0.398&$-0.007$&\dpow{1.048}{-1}&1.177&$-0.636$&\dpow{8.877}{-2}\\
Fe/An&0.163&$-0.002$&\dpow{3.509}{-2}&0.216&$-0.233$&\dpow{5.466}{-2}\\
\multicolumn{7}{l}{Constrained arc cotangent}\\
Dunite&2.668&$-0.209$&\dpow{2.927}{-1}&11.175&$-1.210$&\dpow{2.060}{-1}\\
&0.956&$-0.018$&\dpow{2.730}{-1}&37.747&$-1.567$&\dpow{4.498}{-2}\\
&1.367&$-0.028$&\dpow{4.517}{-1}&10.428&$-1.164$&\dpow{1.803}{-1}\\
Anorth.&1.391&$-0.030$&\dpow{4.267}{-1}&6.537&$-0.914$&\dpow{9.501}{-2}\\
Fe/An&0.978&$-0.018$&\dpow{2.343}{-1}&3.022&$-0.668$&\dpow{3.982}{-2}\\\midrule
\end{tabular}
\end{center}
\end{table}
\begin{table}[t]
\ContinuedFloat%
\caption{(continued)}
\begin{center}
\begin{tabular}{lcccccc}\midrule
&\multicolumn{3}{c}{linear fit (eq. \ref{eqs:parfit}a)}&\multicolumn{3}{c}{logarithmic fit (eq. \ref{eqs:parfit}c)}\\
Data&$\alpha$&$\beta$&misfit&$\alpha$&$\beta$&misfit\\\midrule
\multicolumn{7}{c}{$n$}\\
\multicolumn{7}{l}{Inverse-$r$}\\
Dunite&0.670&0.115&\dpow{9.785}{-2}&0.101&1.664&\dpow{1.274}{-1}\\
&0.979&0.044&\dpow{4.910}{-1}&$-1.913$&3.045&\dpow{2.006}{-1}\\
&1.151&0.041&\dpow{2.126}{-1}&$-0.203$&1.954&\dpow{2.005}{-1}\\
Fe/An&1.499&0.035&\dpow{3.970}{-1}&0.339&1.631&\dpow{9.557}{-2}\\
\multicolumn{7}{l}{Constrained inverse-$r$}\\
Dunite&0.767&0.046&\dpow{9.818}{-2}&0.487&0.730&\dpow{9.288}{-2}\\
&0.445&0.061&\dpow{3.287}{-1}&$-3.181$&3.958&\dpow{4.572}{-2}\\
&0.702&0.055&\dpow{1.421}{-1}&$-1.057$&2.620&\dpow{3.145}{-1}\\
Fe/An&0.812&0.028&\dpow{2.974}{-1}&$-0.098$&1.284&\dpow{5.949}{-2}\\
\multicolumn{7}{l}{Arc cotangent}\\
Dunite&0.654&0.091&\dpow{5.815}{-2}&0.195&1.326&\dpow{8.291}{-2}\\
&0.759&0.038&\dpow{4.110}{-1}&$-1.707$&2.597&\dpow{1.637}{-1}\\
&1.021&0.032&\dpow{1.602}{-1}&$-0.041$&1.539&\dpow{1.731}{-1}\\
Fe/An&1.279&0.027&\dpow{2.885}{-1}&0.396&1.247&\dpow{5.748}{-2}\\
\multicolumn{7}{l}{Constrained arc cotangent}\\
Dunite&0.718&0.033&\dpow{1.025}{-1}&0.511&0.529&\dpow{9.872}{-2}\\
&0.335&0.051&\dpow{2.643}{-1}&$-2.778$&3.360&\dpow{7.146}{-2}\\
&0.624&0.045&\dpow{1.270}{-1}&$-0.754$&2.044&\dpow{2.978}{-1}\\
Fe/An&0.671&0.023&\dpow{2.463}{-1}&$-0.084$&1.066&\dpow{4.870}{-2}\\\bottomrule
\end{tabular}
\end{center}
\end{table}
It is not always obvious whether the linear or the non-linear parameter fits perform better, but by and large, the non-linear formulae eqs.~\ref{eqs:parfit}b and c tend to yield slightly better results and are therefore generally preferred. The more promising fitting models for $p(r)$, i.e., the inverse-$r$ and the arc cotangent models, can be combined with these parameter fits into a tentative general expression in which the corresponding coefficients $a$, $b$, and $n$ in eqs.~\ref{eqs:genform}a and c and \ref{eqs:cform}c are functions of $v$ of the form eqs.~\ref{eqs:parfit}b and c, respectively. These fits are shown in Fig.~\ref{fig:fit1g}, along with the general fitting formulae by \citet{MoAr-Ha16} and by \citet{Pier:etal97} (in modified form).

\section{Discussion}
On a general note, all datasets indicate that the numerical models approach the impedance-match solution at $r\to 0$ the better the higher the velocity of the impactor is; this was already noticed by \citet{AhOKe77}, who suggested that it may be a numerical effect, namely a consequence of the shorter timesteps in models with greater $v$. Another conspicuous feature is the fact that the transition between the near and the far field seems to become sharper and the far-field slope becomes steeper as $v$ increases; this fact is reflected by the $v$ dependence of the exponent $n$ in the model functions.\par
Given the fewer fitting parameters of the constrained fitting model functions eqs.~\ref{eqs:cform}, one would expect them to yield poorer fits to the data than their unconstrained counterparts. Indeed, inspection of the curves and the misfits indicates that this tends to be true, in particular at low $v$, but several fits with the constrained inverse-$r$ or arc cotangent models perform remarkably well, especially at higher $v$. The best fits are usually achieved with the arc cotangent model, with the inverse-$r$ model coming close, whereas the inverse polynomial models give good results only at $v\leq 10$\,km/s. The following discussion will therefore mostly focus on the arc cotangent and the inverse-$r$ models; due to the marked deterioration at high $v$, the inverse polynomial models are not considered further as general models, although their performance for dunite-on-dunite and iron-on-anorthosite impacts at low $v$ is on a par with the other models.\par
The fits to dunite data at three different $v$ spanning the entire range of $v$ for which data are available (Fig.~\ref{fig:fit1}) show the strengths and weaknesses of the different fitting models. The constrained inverse-$r$ and arc cotangent models fit the data from \citet{MoAr-Ha16} almost as well as their own three-part fit and have the advantage of approaching a finite value below the impedance-match solution, whereas the near-field fit of \citet{MoAr-Ha16} would need to be capped to prevent it from growing to infinity as $r\to 0$. The formula by \citet{Pier:etal97} can obviously not be used at these low $v$ at all, because the exponent $n$ changes sign at about $v=5.07$\,km/s, as has already been observed by \citet{JHRoBa12} and \citet{MoAr-Ha16}; it had been calibrated with data for $v\geq 10$\,km/s and should not be used for extrapolation to much lower $v$, just as the formulae by \citet{MoAr-Ha16} were calibrated with and optimized for low-$v$ data and are not suited for extrapolation to high $v$, as can be seen in the plot for $v=60$\,km/s. This deterioration has to be kept in mind when using the \citet{Pier:etal97} formula for planets like Mars, where the (effective) impact velocities are expected to lie below the calibration range of the formula, because it translates into an overestimate of the shocked and heated region. At 10\,km/s, where the datasets from \citet{Pier:etal97} and \citet{MoAr-Ha16} can be combined, all formulae perform at least decently. The \citet{Pier:etal97} fit would be better in the near field if the innermost data points had been used instead of the impedance-match solution, but it is clear that in the far-field, their fit fails to reproduce the steeper decline revealed by the data from \citet{MoAr-Ha16} and returns pressures several times too high. By contrast, the innermost data points from \citet{MoAr-Ha16} suggest a flatter trend than the data by \citet{Pier:etal97} in the near field, and their near-field formula, which is apparently based on only two or three points, does not match the older data well. The fitting models of this study reproduce the data quite well for $r/R\lesssim 10$; beyond that distance, the pressures returned by the inverse-$r$ and arc cotangent models begin to lie a bit higher than the data points, whereas the inverse polynomial model continues to give a quite good fit, although its slope may become too steep at distances greater than those sampled by the models. The reason for the deterioration may lie in the fact that the two datasets, although agreeing quite well, introduce conflicting trends in the crucial transition region. I felt unable to decide which data are to be preferred at different $r$ but would expect the proposed model functions to work better if one single, consistent dataset for the entire $r$ range were available. This highlights that a good sampling of the near field and the far field as well as the transitional region is crucial to achieve a fit that works well at all $r$, and the problem is that especially at low to intermediate $v$, such good sampling is generally not achieved, probably due to the computational cost. Nonetheless, the individual fits for each $v$ give good approximations at all $v$, especially for the preferred inverse-$r$ and arc cotangent functions, as can be seen in Fig.~\ref{fig:fit1} and Table~\ref{tab:fit1}; even for the 10\,km/s fit, they lie mostly within the misfit range, especially in the critically important intermediate distance range where the shock temperature approaches the solidus (cf. Supplementary Material, pp.~27--28). They should therefore serve their purpose well.\par
The apparently good performance of some of the model functions for $p(r)$ at all $v$ motivated the construction of $v$-dependent parameters to be used with the $p$ model. The results as applied to the same dunite datasets are shown in Fig.~\ref{fig:fit1g} together with the general fits from \citet{MoAr-Ha16} and \citet{Pier:etal97}. At $v$ up to 10\,km/s, the \citet{MoAr-Ha16} formula with $v$-dependent coefficients works best, whereas the \citet{Pier:etal97} formula cannot be used or has problems in the far field. Some of the model functions from this study show promise, but they do not perform quite as well as one might have hoped; especially the constrained inverse-$r$ model disappoints irrespective of the form chosen for $n(v)$ and is therefore not considered further for the general fit. A possible reason can be identified by analyzing the parameter fits in Fig.~\ref{fig:parfit}. Although a general overall trend can be clearly identified for all three parameters and justifies the application of one function for the entire $v$ range, the two datasets also show distinct individual trends. Although it has been tried to link them at 10\,km/s by using the parameter values for the combined fit, it becomes clear that the match between the two datasets is not perfect. Their different trends are probably a consequence of their different strengths in terms of spatial sampling. Again, a dataset produced from a single model suite with good spatial sampling in all fields would likely make a better fit possible. Altogether, the general $v$-dependent fit performs clearly less well than the individual fits even for the preferred model functions, and the fitting curves lie outside the misfit range in several instances, albeit often only marginally so (cf. Supplementary Material, pp.~29--30); when judging the quality of the fit, one should keep in mind that the distortion due to the logarithmic scale tends to make discrepancies look worse than they are in the far field, where the shock pressure has already dropped to relatively small values. However, the quality of the fit tends to improve towards higher $v$. Depending on the accuracy requirements of the problem at hand, end users may decide to use the general $v$-dependent fit or interpolate between two of the more accurate individual-$v$ fits from Table~\ref{tab:fit1} if their $v$ is not covered by one of the available fits.\par
Given the shortcomings of the available data, it is difficult to decide which of the parameter fitting models in eqs.~\ref{eqs:parfit} are best. In the case of $a(v)$, the power law seems to work a bit better than the linear fit, and it has the benefit that $a(0)=0$ for positive exponents, which is the physically expected behavior for all three pressure decay models, because $a$ sets the pressure scale. The properties of $b(v)$ are less clear for all three decay models, although for physical reasons a function that does not become zero at $v=0$ is preferable for the inverse-$r$ (and also the inverse polynomial) model to avoid that $p\to\infty$ for $r, v\to 0$. Whereas $b(v)$ increases with $v$ in those models, it decreases for the arc cotangent models, if a unique trend can be identified at all, i.e., for a power-law $b(v)$ the exponent is negative, and there is a pole at $v=0$; however, as the arc cotangent tends to zero for increasing arguments, this is not a problem. The logarithmic form of $n(v)$ was proposed by \citet{AhOKe77} and also seems to work well in most of the cases investigated here. As $n$ is always positive in our fits, it seems that a change of sign is not a desirable feature of a functional form for $n(v)$ and would limit the range of $v$ where it can be used; the logarithmic form may thus be problematic as $v\to 0$. On the other hand, very low impact velocities are probably of interest only in rather special circumstances in the context where such formulae are applied, because they approach the speed of sound, and the corresponding collision would not be a hypervelocity impact with the concomitant shock phenomena.\par
In the literature it has been common practice to distinguish between a near field and a far field, and possibly some transitional regime, and to draw the boundary between them based on the inspection of the available data. Although at least the division into two domains is clearly justified, it is often not possible to define the position of the boundary precisely; for instance, the fits for the near field and the mid field in the numerical models by \citet{MoAr-Ha16} especially at low $v$ are based on very few ($<5$) points, and the regression lines do not cross at the nominal boundaries. Appropriate coverage of both the near field and the far field is crucial, because insufficient sampling of the near field and the transitional region can easily cause the fit to yield poor values for $r=0$ and give a wrong impression about the validity of the impedance-match solution. The problem seems to be more salient at low $v$, therefore the fits to the low-$v$ data for dunite from \citet{MoAr-Ha16} and for the 5\,km/s data from \citet{AhOKe77} must be interpreted with caution in this respect; on the other hand, the high-velocity data often do not cover the far-field to distances as large as those from the low-$v$ \citet{MoAr-Ha16} data, and so the latter can be expected to provide more reliable constraints for the far-field decay. Ideally, coverage would always be as good as in the combined datasets for dunite at $v=10$\,km/s.\par
A reliable fit especially in the near field is not only desirable as a ``final product'' for the impact modelers, but it is also important for users of that final product from other areas such as mantle convection modeling. For instance, the shock heating that produces a thermal anomaly and possibly melt in the central area is a function of the shock pressure. The waste heat produced by the unloading of the shocked material has been estimated by \citet{GaHe63} and was reformulated by \citet{WAWatt:etal09} as a function of shock pressure as
\begin{equation}
\Delta H=\frac{\Delta p}{2\varrho S}\left(1-\frac{1}{\Phi}\right)-\left(\frac{c}{S}\right)^2 (\Phi-\ln\Phi-1)\label{eq:dT}
\end{equation}
with
\addtocounter{equation}{-1}\begin{subequations}
\begin{align}
\Phi&=-\frac{2 S \Delta p}{c^2 \varrho} \left(1-\sqrt{\frac{4S\Delta p}{c^2 \varrho}+1}\right)^{-1}\\
\Delta p&=p-p_\mathrm{l},
\end{align}
\end{subequations}
where $p_\mathrm{l}$ is the lithostatic pressure, $\varrho$ is the density, $c$ is the bulk sound speed, and $S$ is a parameter related to the Hugoniot curve (cf. \ref{app:im}). This relation can be divided by the isobaric heat capacity $c_p$ to yield the temperature increase, which, after correcting for latent heat consumption upon melting and/or vaporization, is commonly used for calculating the post-impact temperature field in large-scale mantle convection models that consider the effects of impacts \citep[e.g.,][]{Rees:etal02,WAWatt:etal09,JHRoAr-Ha12}. Hence, a poor fit may result in poor estimates of the magnitude of the temperature anomaly and its spatial extent, including the melt-producing region; even if the temperature is capped at or near the solidus in such models for technical reasons, the size of the shocked region remains a concern.\par
Figure~\ref{fig:dTg} shows the thermal effect for the pressures from Fig.~\ref{fig:fit1g}. The temperatures in the central part of the target region are of course much higher than can realistically be expected and are to be taken as measures of the waste energy rather than as actual temperatures, because effects like latent heat consumption in various phase transitions are not included here. The parts worth more quantitative consideration are those at lower temperatures, i.e., where $\Delta T\lesssim 1000$\,K, because they mark the boundaries of the partially molten region; the actual maximum $\Delta T$ of interest in this regard depends on the pre-impact temperature and will be higher for shallow depths that affect only the crust than for giant impacts that penetrate into the already hot uppermost mantle. As can be seen from these plots, for low to intermediate $v$ the use of different pressure decay models translates into temperature differences of several hundred kelvin in the far field or to differences of several times the impactor radius for the size of the heated and molten region, which corresponds to tens to hundreds of kilometers for giant impacts.\par
\begin{figure}[b]
\caption{Waste heat and temperature increase (before accounting for melting) due to shock compression in the dunite-on-dunite models as calculated from eq.~\ref{eq:dT} with the pressures plotted in Fig.~\ref{fig:fit1g}. A specific heat of 1250\,J/(kg\,K) was assumed, and the lithostatic pressure was set to zero for simplicity. The actual temperature after accounting for latent heat of fusion may be much lower in the supersolidus regions of the model.\label{fig:dTg}}
\end{figure}
\begin{figure}[!p]
\begin{center}
\includegraphics[width=0.83\textwidth]{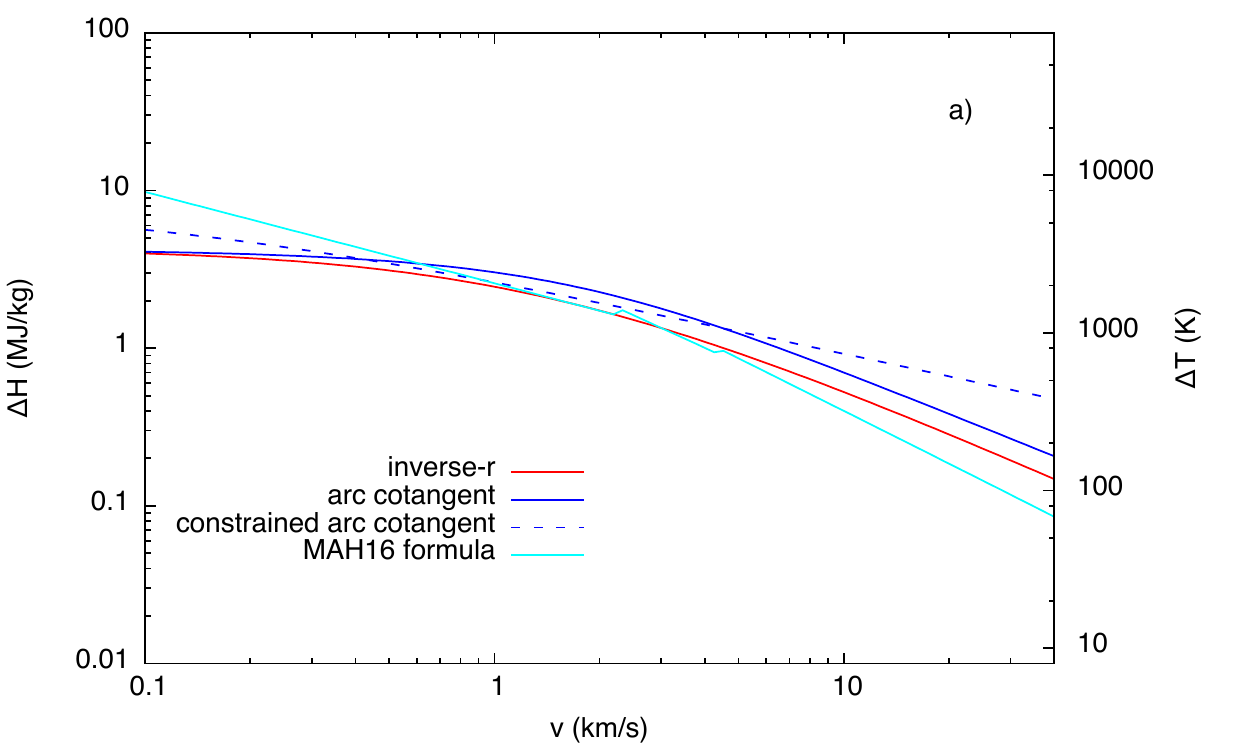}
\includegraphics[width=0.83\textwidth]{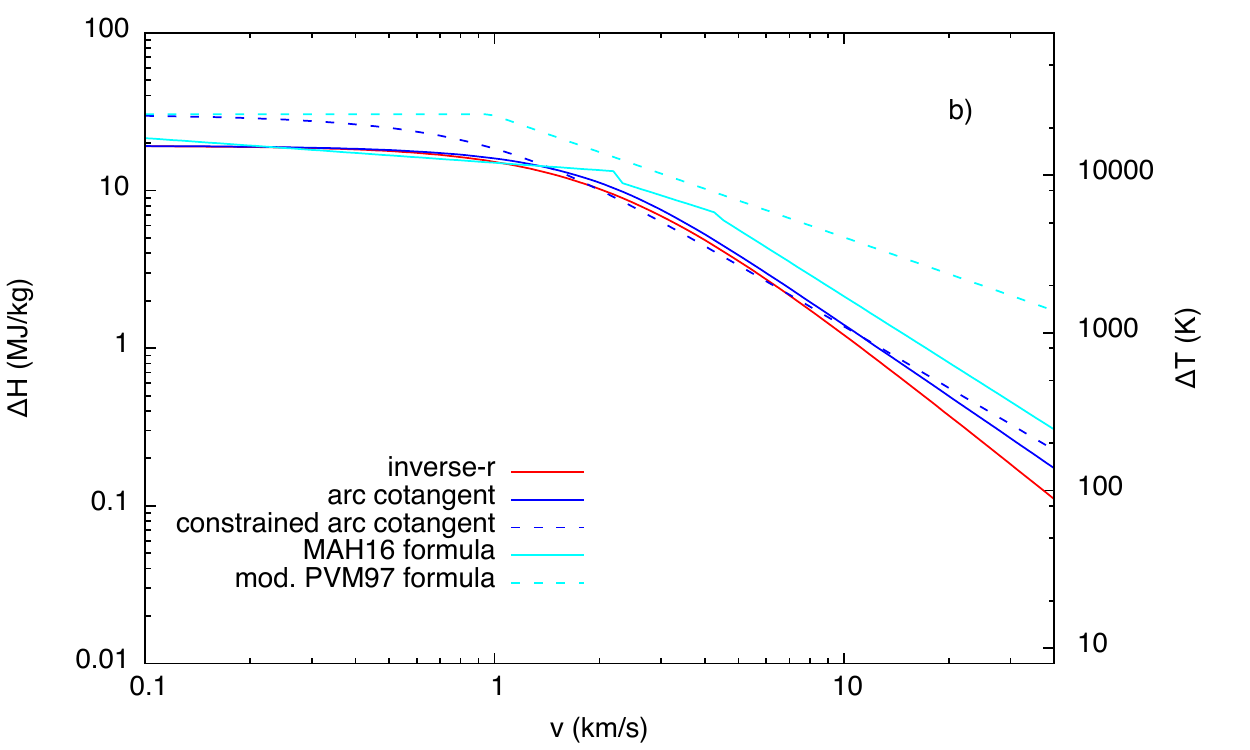}
\includegraphics[width=0.83\textwidth]{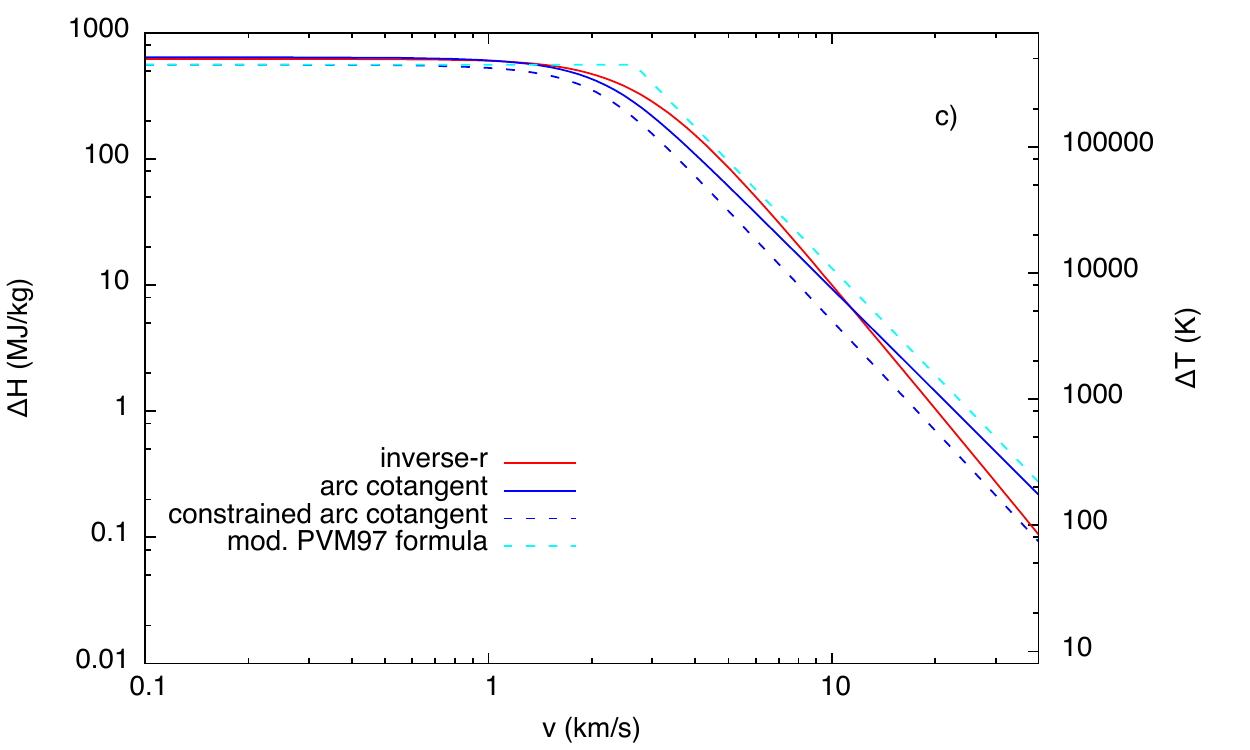}
\end{center}
\end{figure}
The use of a single, continuous, and smooth function does not only make it unnecessary to define explicitly such domains as a near field and a far field and to decide upon boundaries for the purpose of fitting, but in principle offers the opportunity to derive the position of a boundary between a near field and a far field on the basis of well-defined properties of the fitting function; some basic properties of the inverse-$r$ model and the two arc cotangent models, which will be needed below, are given in \ref{app:derivs}.\par
Keeping in mind the traditional notion that the near field is characterized by a near-constancy of the pressure, the most straightforward possibility is to define a certain fraction $\varphi$ of the maximum pressure as the boundary, so that for the inverse-$r$ and the arc cotangent models the boundary would simply lie at
\begin{subequations}\label{eqs:rphi}
\begin{align}
r_\varphi&=R\sqrt[n]{\frac{a}{\varphi p_\mathrm{max}}-b}=R\sqrt[n]{\frac{1-\varphi}{\varphi}b},\\
r_\varphi&=R\sqrt[n]{\frac{\cot\frac{\varphi p_\mathrm{max}}{a}}{b}}=R\sqrt[n]{\frac{\cot\frac{\varphi\pi}{2}}{b}}
\end{align}
\end{subequations}
respectively. However, inspection of the data and the curves suggests that a useful definition of the boundary should relate it to the curvature, whereby it must be kept in mind that the usual logarithmic representation distorts the curves and may give a partly misleading visual impression of the position of such key points. If the boundary is to be defined as the inflexion point, it may be located at
\begin{subequations}\label{eqs:infl}
\begin{align}
r_\mathrm{infl}&=R\sqrt[n]{\frac{n-1}{n+1}b}\\
r_\mathrm{infl}&=R\sqrt[2n]{\frac{n-1}{b^2(n+1)}}
\end{align}
\end{subequations}
for the inverse-$r$ and the arc cotangent models, respectively, under the condition that $n\geq 1$. As $b$ and $n$ are functions of $v$, so is the inflexion point. Table~\ref{tab:fit1} shows that the individual fits with the unconstrained model functions fulfill this requirement in most cases, with $n$ usually lying between 1 and 3, but in constrained models, including the constrained arc cotangent model, $n$ is often smaller than 1. In this case, there is no inflexion point, and the pressure curve has a cusp instead of a flat maximum at $r=0$.\par
Alternatively, the point of strongest negative curvature could be considered for defining the boundary. The curvature of a function $f(r)$ is defined by
\begin{subequations}
\begin{equation}
K=\frac{f''}{\left(1+f'^2\right)^\frac{3}{2}},
\end{equation}
where primes indicate derivatives with respect to $r$. Hence, the task is to find the global minimum of $K$, which is a root of its first derivative
\begin{equation}
\frac{\df K}{\df r}=\frac{f'''+(f'''-3f')f''^2}{\left(1+f'^2\right)^\frac{5}{2}}
\end{equation}
\end{subequations}
at which the second derivative is positive. Although we know that it lies between the inflexion point and the maximum at $r=0$, there is unfortunately no analytic solution for the root for any of the functions proposed here, but the derivatives can be determined so that the root can be found numerically. However, its usefulness is limited by the fact that it coincides with the maximum if $n\leq 2$ (cf. \ref{app:derivs}), which is often the case for velocities up to 10--30\,km/s.\par
\begin{figure}
\begin{center}
\includegraphics[width=\textwidth]{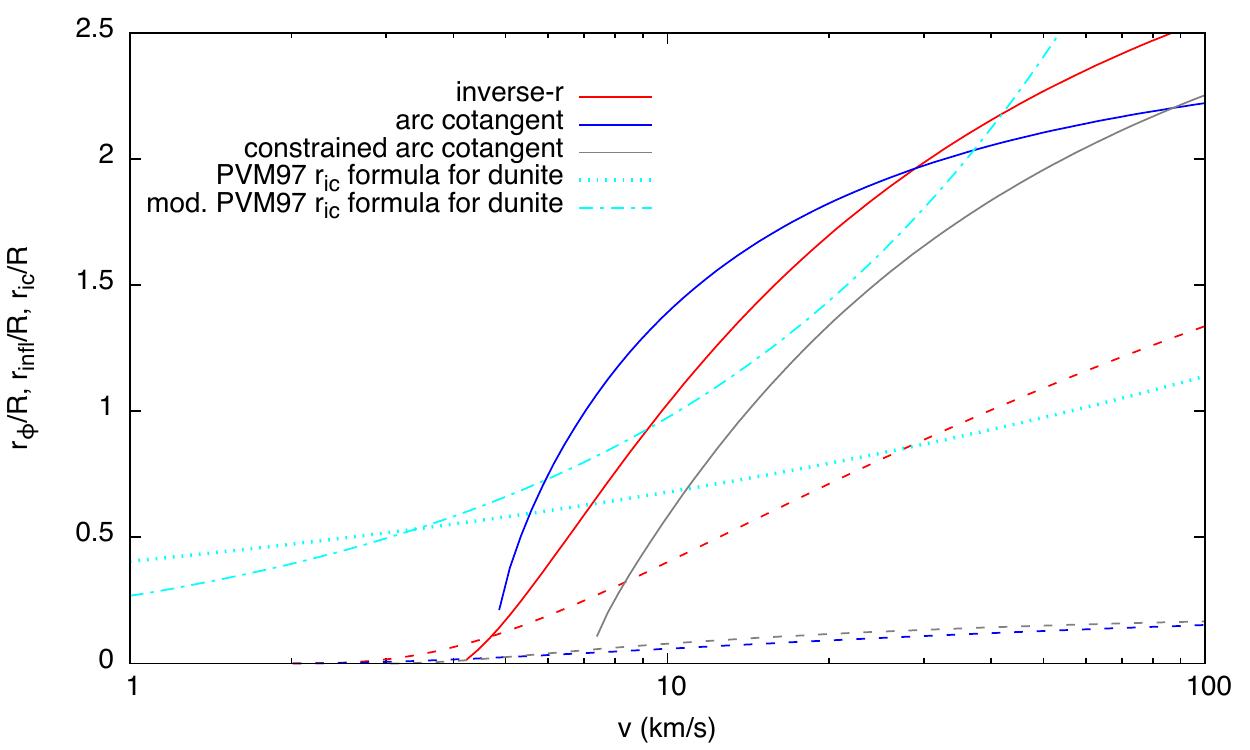}
\end{center}
\caption{Normalized distance of the inflexion point (solid lines) and of the point at which $p$ has dropped to $\varphi=0.95$ of its peak value (dashed lines) as a function of impact velocity $v$ for dunite, according to eqs.~\ref{eqs:rphi} and \ref{eqs:infl}, respectively. $b$ has the form of eq.~\ref{eqs:parfit}b, and $n$ follows eq.~\ref{eqs:parfit}c. For comparison, the formulae for the radius of the isobaric core, $r_\mathrm{ic}$, from \citet{Pier:etal97} and its modified form are also given, with their parameters for dunite.\label{fig:n2f}}
\end{figure}

Fig.~\ref{fig:n2f} shows $r_\mathrm{infl}$ and $r_\varphi$ for $\varphi=0.95$ as functions of $v$ along with the formula by \citet{Pier:etal97} in original and modified form for the radius of the isobaric core,
\begin{equation}
\lg\frac{r_\mathrm{ic}}{R}=a_\mathrm{ic}+b_\mathrm{ic}\lg v \Leftrightarrow
\frac{r_\mathrm{ic}}{R}=a'_\mathrm{ic} v^{b_\mathrm{ic}},
\end{equation}
with $v$ in km/s; the values used here are not their general values, but their values for dunite from their Table~II, whereby their erroneous value 0.022 for $b_\mathrm{ic}$ is replaced by 0.22. The original coefficients are thus $a_\mathrm{ic}=-0.391$ ($a'_\mathrm{ic}=10^{a_\mathrm{ic}}=0.406$) and $b_\mathrm{ic}=0.22$, the coefficients for the modified form are $a_\mathrm{ic}=-0.5718$ ($a'_\mathrm{ic}=0.268$) and $b_\mathrm{ic}=0.561$. Given the insufficiencies in the $v$-dependence fits of the $a$, $b$, and $n$, the curves should only be seen as a crude guide and are not suited for extrapolation to lower $v$. As already mentioned, the point of maximum curvature detaches itself from the maximum only at higher $v$ for dunite, whereas the fit for the anorthosite target models from \citet{AhOKe77} suggests that the separation takes place at lower $v$ already in that case. This would point to a material dependence, which is plausible, but given the scatter in the \citet{AhOKe77} data, it could also be an artifact. It may be tempting to infer a change in dynamic behavior from the existence and position of the point of maximum curvature and/or the inflexion point, but given that the model functions are not derived from physical principles but are ad hoc constructs only designed to fulfill certain physical requirements, such inferences would probably be an overinterpretation.\par
At this point, some remarks on the formula by \citet{Pier:etal97} with its original coefficients are necessary, because this formula has been used by several studies. The application of their original formula (but already with corrected coefficient $b_\mathrm{ic}$) to their own data for $v=10$ and 60\,km/s is shown in Fig.~\ref{fig:fit1}b and c and reveals that their equation gives an increasingly bad description of $r_\mathrm{ic}$ toward higher $v$; of the four curves available from their paper, only the result for $v=10$\,km/s was found to be satisfactory. This problem prompted an attempt to determine the coefficients at least for dunite anew following their method, the result of which led to the coefficients of the modified version. \citet{Pier:etal97} explained, with reference to their Figure~4 that they ``estimated the size of the isobaric core by determining the point where the two pressure decay regimes intersect''. These regimes were defined ``by fitting a line through the two tracer points closest to the impact point for the first regime and by a least-squares fit of a number of tracer points for the second (avoiding the intermediate region, where the shift from one regime to the other occurs).'' The expression ``fitting a line through [\,\dots] two tracer points'' can be understood in two ways: first, as drawing a line going through these two points and beyond, which however should not be called a ``fit''; and second, as taking the mean of the two values as a constant, which is not what is usually meant when speaking of ``fitting'', but would result in an isobaric region sensu stricto. The first method has the disadvantage of being very sensitive to errors in the values of these two points, which may easily translate into substantial shifts in the position of the intersection; moreover, the fitted line may never intersect the far-field fit at all. I have therefore tried both options and decided to use the second, ``isobaric'' for the modified formula; it gives the smallest possible radius under the given conditions. For the far-field fit, it is not always clear how many points should be included, as \citet{Pier:etal97} do not (and actually cannot) give a specific number. In order to reproduce their $r_\mathrm{ic}$ value for $v=10$\,km/s, which can be seen in their Figure~8a, one has to include almost all of the remaining points, effectively including the intermediate region they tried to exclude; indeed, they hardly seem to reach the far field proper in the first place, as the far-field data from \citet{MoAr-Ha16} reveal, so that the satisfactory performance of their formula applies only to the intermediate region. Using the far-field fit from \citet{MoAr-Ha16} at this $v$ instead did not give a better overall result, however, and was therefore dismissed. The $p$ curve for $v=20$\,km/s from \citet{Pier:etal97} features an odd kink at $r/R\approx 4$, where the curve flattens again and which is not seen in that form in the other curves and is probably an artifact. If one uses the outermost points for the far-field fit, the result has too shallow a slope and will intersect the near-field line too closely to the origin. Their value for $v=20$\,km/s as plotted in their Figure~8a indicates that this is what they did, but a better result would probably have been obtained by using, or at least including, the points at $2\lesssim r/R\lesssim 4$; I tried all three possibilities and favor the fit only with points between $r/R\approx 2$ and the kink, which gives the largest $r_\mathrm{ic}$. In the fit for $v=60$\,km/s, two sets of points, $r/R\gtrsim 4$ and $r/R\gtrsim 6$, may be considered, but the difference is minor; eventually the latter was used. The final modified formula gives results that are worse at $v=10$\,km/s, but substantially better at all higher $v$. In any case, a general problem with this parameterization that neglects the intermediate region is that the region that is strongly shock-heated to temperatures even above the melting point is artificially augmented in the modified version, whereas the original form results in an underestimate at higher $v$. This can introduce substantial errors in terms of the thermal effect and melt production especially in the case of giant impacts.\par
\begin{figure}
\begin{center}
\includegraphics[width=\textwidth]{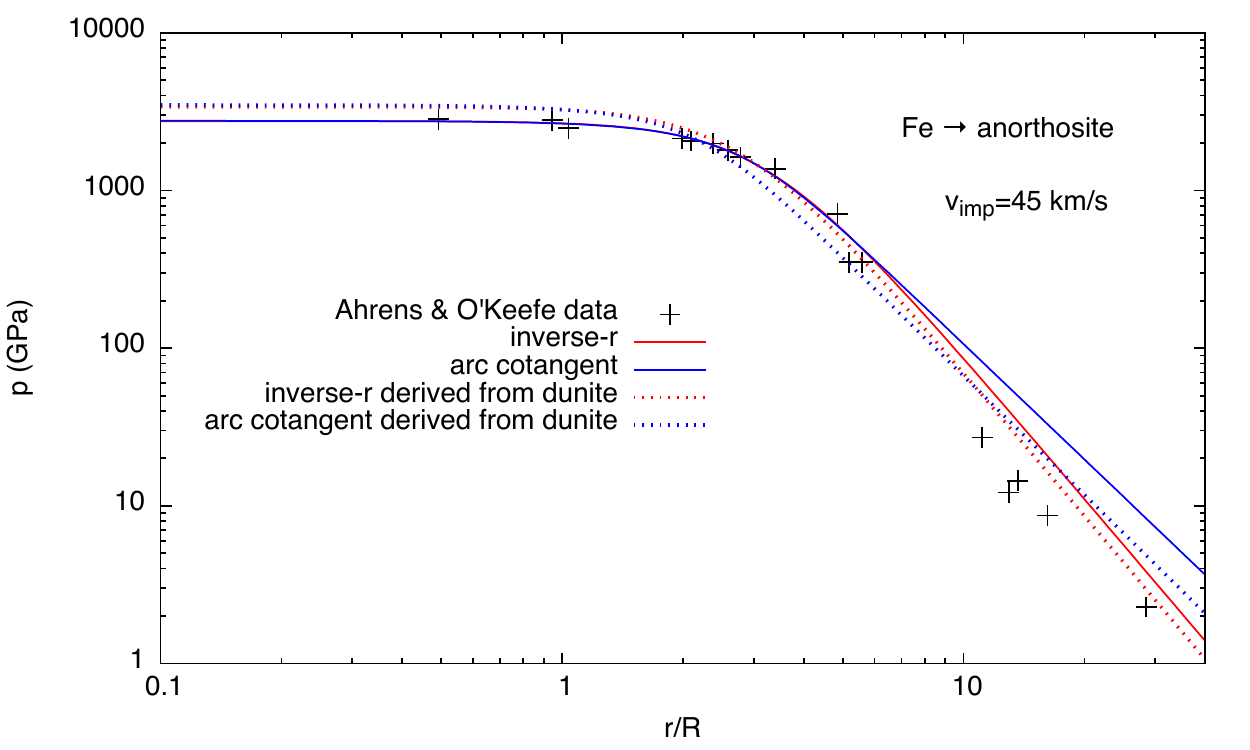}
\end{center}
\caption{Shock pressure for impacts of an iron projectile on an anorthosite target at a velocity of 45\,km/s. The points are data from the numerical study by \citet{AhOKe77}, and the solid lines are the inverse-$r$ and arc cotangent fits to these data according to Tab.~\ref{tab:fit1}. The dotted lines are the inverse-$r$ and arc cotangent parameterizations from the dunite data, calculated with parameters for $v=45$\,km/s with eqs.~\ref{eqs:parfit}b and c, and rescaled with the impedance-match solution for an iron-on-anorthosite impact.\label{fig:extr}}
\end{figure}

Another question of interest is whether the pressure decay for an arbitrary material combination for projectile and target can be predicted if the decay law for a given combination is known. The few datasets available (cf. Fig.~\ref{fig:data}) suggest that in normalized distance and pressure coordinates, the data for different materials indeed follow fairly similar trends at a given $v$, although the scatter in the \citet{AhOKe77} data introduces considerable uncertainty; nonetheless, similar conclusions had already been reached by the original authors of the respective papers, prompting \citet{Pier:etal97} to formulate their relation for the decay exponent for all materials investigated (except ice) in a uniform way. In the framework of this work, it means that it may be possible to arrive at a decent result for an impact of a projectile made of material A into a target of material A (or B) by using an established fit for material C in normalized coordinates and rescaling it with the impedance-match solution for materials A and B, which can be determined independently from the material properties. An example is shown in Fig.~\ref{fig:extr}, where the iron-on-anorthosite impact data from \citet{AhOKe77} at $v=45$\,km/s are shown with their best fits according to Tab.~\ref{tab:fit1} and with a prediction derived from the dunite-on-dunite impact data.\par
The improvement of the phenomenological description of pressure decay notwithstanding, it must be kept in mind that analytical formulae of the type considered here are only idealized, approximate descriptions of some fundamental characteristics of an impact process. As such, they may be suited for use in settings where a first-order description is sufficient and basic characteristics need to be estimated quickly and efficiently, for instance in global mantle convection models that cannot resolve much detail of an impact anyway. In such models, which are the incentive for this study, the most important task is to obtain a reasonable estimate of the spatial extent and amplitude of the temperature anomaly for one or several successive impacts; in such a context, the impact velocity from the formula will usually be replaced by its vertical component, assuming an impact angle of 45\textdegree, and effects related to the free surface such as destructive interference of the direct shock wave and its reflection from the surface may be dealt with by ``post-processing'' the calculated shock pressure according to an approach developed by \citet{Melosh84}. The essential end result of the parameterized shock pressure formula is a strong thermal anomaly of a certain size that will usually be located at some depth in the upper mantle of a planet and modify the regional or even the global convection patterns, even though it is common practice for technical reasons to cap the thermal anomaly at a temperature close to the solidus \citep[e.g.,][]{Rees:etal04,WAWatt:etal09,JHRoAr-Ha12}. If such models also include the compositional effects of melting, the thermal anomaly generated by the shock induces a compositional anomaly that entails dynamical effects of its own \citep[e.g.,][and submitted]{Rees:etal04,Breu:etal16,RuBr16}. The multitude of details found in a real impact, much of it on length scales below the resolution of global convection models, is necessarily neglected, and much of it would be destroyed by stirring in the dynamically vigorous aftermath of the impact anyway. Accounting for the departure from axisymmetry in oblique impacts or a more accurate description of surface effects in an analytical model would be welcome further improvements, but they are likely to complicate the expressions considerably, if they are possible at all.

\section{Conclusions and Outlook}
The parameterized expressions for the shock pressure in an impact commonly used in situations where reasonable estimates must be obtained efficiently have a form that requires an artificial subdivision of the affected volume into different decay regimes and additional assumptions about the maximum pressure. In order to avoid some of these shortcomings, it is suggested to replace the conventional formulations with a type of function that is smooth in the entire domain and fulfills certain physical requirements like boundedness by design. Specifically, a form as given in eqs.~\ref{eqs:genform}a or c, i.e., the inverse-$r$ or the arc cotangent model, are found to be useful, and fits have been carried out for dunite-on-dunite impacts at velocities between 4 and 60\,km/s and for anorthosite or iron-on-anorthosite impacts between 5 and 45\,km/s on the basis of published numerical models.\par
A tentative parameterization has also been developed in which the fitting parameters that control the shape of the curve are functions of the impact velocity, but is found to give satisfactory results only at large $v$, as a consequence of the limitations in sampling density, the extent of the modeling domain in the numerical models, but also the inhomogeneities in, or slight inconsistencies between, the available datasets. Users may therefore still prefer to interpolate linearly between two sets of coefficients from Table~\ref{tab:fit1} bracketing the target $v$ or devise their own fits for eqs.~\ref{eqs:parfit} to a suitable subset of the coefficients from that table. It seems that a non-linear functional form for the dependence of the fitting parameters on $v$ is preferable, but in order to arrive at a definitive decision, a comprehensive survey covering a wide range of impact velocities (from a few to several tens of km/s) and a broader choice of materials is needed. Such a survey should also establish more firmly the tentative observation from this and earlier studies that the decay law is, at least within the limits of certain material classes, independent from the material in normalized distance--pressure space. The application of the decay model functions proposed here to the data of this survey will crucially depend on sufficiently dense sampling at all distances, i.e., even in the very near field and in the region of strongest curvature. Similar model functions can also be applied to the particle velocity.\par
In studies that use a parameterization of pressure decay such as mantle convection models that include large impacts, an accurate representation of the impact-induced temperature anomaly is of great interest with respect to the subsequent dynamics of the mantle and the production of melt. It is for applications of this sort that the model functions proposed here have been developed. It is hoped that they will be improved further, for instance by taking into account the dependence from impact angle, with data from new, dedicated numerical models and replace the hitherto used function developed by \citet{Pier:etal97}, which was found to be problematic in this respect.

\section*{Acknowledgments}
I thank Julien Monteux for some further explanations of his paper. The comments of James Roberts, Ross Potter, and an anonymous reviewer helped to improve the paper, in particular to clarify the purpose and applicability of parameterizations of the type discussed here. This work was funded by grant Ru~1839/1-1 from the German Science Foundation (DFG), with additional initial funding from the Helmholtz Alliance project ``Planetary evolution and life''.

\appendix
\section{Impedance-match solution}\label{app:im}
The impedance-match solution for the peak pressure generated in the collision of two infinite colliding planes has been given by various authors. This brief summary essentially follows \citet{Melosh89,Melosh11}.\par
The particle velocity $u$ in the target is given by
\begin{equation}
u=
\begin{cases}
\frac{-\mathcal{B}+\sqrt{\mathcal{B}^2-4\mathcal{A}\mathcal{C}}}{2\mathcal{A}},&\text{different materials},\\
-\frac{\mathcal{C}}{\mathcal{B}}=\frac{v}{2},&\text{same materials}
\end{cases}\label{eq:uic}
\end{equation}
with
\addtocounter{equation}{-1}\begin{subequations}\begin{align}
\mathcal{A}&=\varrho S-\varrho_\mathrm{imp} S_\mathrm{imp}\\
\mathcal{B}&=\varrho c+\varrho_\mathrm{imp}(c_\mathrm{imp}+2 S_\mathrm{imp} v)\\
\mathcal{C}&=-\varrho_\mathrm{imp}v_{\mathrm{imp}}(c_\mathrm{imp}+S_\mathrm{imp} v)
\end{align}
\end{subequations}
\citep[e.g.,][eqs.~6.5,6.6]{Melosh11}; the properties used here are those of the unshocked material. $c$ is the bulk sound speed of the material. $S$ corresponds to the slope in $U$--$u$ space of the Hugoniot curve of a shock front propagating with velocity $U$. It is related to the zero-pressure Grüneisen parameter $\gamma_0$ by
\begin{subequations}
\begin{equation}
S=\frac{1+\gamma_0}{2}
\end{equation}
and to the parameters commonly designated as $A$, $B$, $a$, and $b$ of the Tillotson equation of state by
\begin{equation}
S=\frac{1}{2}\left(1+\frac{B}{A}+\frac{a+b}{2}\right)\label{eq:tilleos}
\end{equation}
\citep[eqs.~AII.7.4]{Melosh89}. The shock pressure in the target is then
\end{subequations}
\begin{equation}
p=\varrho(c+Su)u.
\end{equation}

\section{Properties of the model functions}\label{app:derivs}
Some basic properties of the three model functions of interest are summarized below, with emphasis on $n\geq 1$, $a>0$, and $r\geq 0$, the latter two conditions being imposed by the physical requirement of having a positive pressure and by the radial symmetry of the problem. For the purpose of this discussion, it is more convenient to replace the variable $r$ by $x=r/R$.

\subsection{Inverse-$r$ model}
The first three derivatives of the inverse-$r$ model function
\begin{equation}
p(x)=\frac{a}{b+x^n}
\end{equation}
are:
\begin{subequations}
\begin{align}
p'=\frac{\df p}{\df x}&=-an\frac{x^{n-1}}{\left(b+x^n\right)^2}\\
p''=\frac{\df^2 p}{\df x^2}&=an\frac{(n+1)x^n-(n-1)b}{\left(b+x^n\right)^3}x^{n-2}\\
p'''=\frac{\df^3 p}{\df x^3}&=-an\frac{(n+1)(n+2)x^{2n}-4(n^2-1)bx^n+(n-1)(n-2)b^2}{\left(b+x^n\right)^4}x^{n-3}.
\end{align}
\end{subequations}
Obviously, $p'(0)=0$, and as $p'<0$ for all $x>0$ for the parameters of interest (except $n=1$), the point at $x=0$ is effectively a global maximum thanks to the symmetry of the problem, fulfilling one of the requirements made at the beginning. However, for $1\leq n<2$, the second derivative, and hence also the curvature, has a pole at $x=0$, which means that in this interval the point of maximum curvature coincides with the maximum at $x=0$; for $n=1$, the maximum of $p(x)$ turns into a cusp. Hence, it is only for $n\geq 2$ that the function will begin to feature a distinct ``flat'' region around the maximum as the point of strongest curvature moves away from the maximum.\par
There is an inflexion point at
\begin{equation}
x_\mathrm{infl}=\sqrt[n]{\frac{n-1}{n+1}b},
\end{equation}
and so the point of maximum curvature must lie between 0 and $x_\mathrm{infl}$.\par
For $(r/R)^n\gg b$, the inverse-$r$ model approaches asymptotically the traditional formulation given in eq.~\ref{eq:tradform} used by previous authors:
\begin{equation}
\frac{a}{b+\left(\frac{r}{R}\right)^n}\approx\frac{a}{\left(\frac{r}{R}\right)^n}
\Leftrightarrow \lg p=\lg a-n\lg\frac{r}{R};
\end{equation}
a similar relation holds also for the inverse polynomial model if $r/R\gg b$.

\subsection{Arc cotangent models}
The first three derivatives of the arc cotangent model function
\begin{equation}
p(x)=a\arccot\left(bx^n\right)
\end{equation}
are:
\begin{subequations}
\begin{align}
p'=\frac{\df p}{\df x}&=-abn\frac{x^{n-1}}{1+b^2x^{2n}}\\
p''=\frac{\df^2 p}{\df x^2}&=abn\frac{(n+1)b^2x^{2n}-(n-1)}{\left(1+b^2x^{2n}\right)^2}x^{n-2}\\
p'''=\frac{\df^3 p}{\df x^3}&=-abn\frac{(n^2+3n+2)b^4x^{4n}-(6n^2-4)b^2x^{2n}+(n^2-3n+2)}{\left(1+b^2x^{2n}\right)^3}x^{n-3}.
\end{align}
\end{subequations}
For the same reasons as for the inverse-$r$ model, the point at $x=0$ is effectively a global maximum for $n>1$ and a cusp for $n=1$, and the same distinction has to be made between the domains $1\leq n<2$ and $n\geq 2$.\par
An inflexion point is located at
\begin{equation}
x_\mathrm{infl}=\sqrt[2n]{\frac{n-1}{b^2(n+1)}}
\end{equation}
and helps to bracket the point of maximum curvature.\par
Euler found the following expression for the inverse cotangent:
\begin{equation}
\arccot z=z\sum\limits_{k=1}^\infty \frac{(2k-2)!!}{(2k-1)!! (z^2+1)^k}
\end{equation}
\citep[e.g.,][]{Wetherfield96}, where !! is the double factorial. Substituting $z=bx^n$ and letting $z\gg 1$ yields
\begin{equation}
a\arccot\left(bx^n\right)\approx a\sum\limits_{k=1}^\infty \frac{(2k-2)!!}{(2k-1)!! (bx^n)^{2k-1}}\approx
\frac{a}{bx^n},
\end{equation}
i.e., to first order and at large distances, the dependence takes the same form as for the other models and the traditional formulation.


\end{document}